\documentclass{article}
\usepackage[utf8]{inputenc}
\usepackage{lscape}
\usepackage[margin=1.0in]{geometry} 
\usepackage{graphicx}
\usepackage{booktabs}
\usepackage{longtable}
\usepackage{tabularx}
\usepackage[skip=0pt, labelfont=bf, font=footnotesize]{caption}
\usepackage{threeparttable}
\usepackage{threeparttablex}
\usepackage[capposition=top]{floatrow}
\usepackage[T1]{fontenc}
\usepackage{etoolbox}
\BeforeBeginEnvironment{tabular}{\small}
\newcounter{magicrownumbers}

\usepackage{nameref}
\usepackage{tikz}
\usepackage[bottom]{footmisc}
\usepackage{authblk}

\usepackage{booktabs} 
\usepackage{nicefrac}  
\usepackage{microtype}
\usepackage[utf8]{inputenc}

\usepackage{wrapfig}
\usepackage{multirow}

\usepackage{amssymb}
\usepackage{amsthm}
\usepackage{bbm}
\usepackage{bbm}
\usepackage{array}
\usepackage[colorlinks=true,urlcolor=purple,
linkcolor=purple,citecolor=purple]{hyperref}
\usepackage{amsmath}
\usepackage{mathtools}
\usepackage{float}

\usepackage{lscape}
\usepackage{graphicx}

\usepackage{booktabs}
\usepackage{longtable}
\usepackage{threeparttable}
\usepackage{threeparttablex}

\usepackage{mathrsfs}
\usepackage{graphicx}
\usepackage{color}
\usepackage{natbib}

\usepackage{algorithm}
\usepackage{algpseudocode}
\usepackage{thmtools,thm-restate} 

\makeatletter
\def\paragraph{\@startsection{paragraph}{4}%
  \z@\z@{-\fontdimen2\font}%
  {\normalfont\bfseries}}
\makeatother

\usepackage{tikz}
\usetikzlibrary{decorations.markings, backgrounds, calc, positioning, shapes, shadows, arrows, fit, automata}
\usepackage{tikz-cd}
\usetikzlibrary{arrows.meta}
\tikzset{>={Latex}}

\newcommand{\beginsupplement}{%
        \setcounter{table}{0}
        \renewcommand{\thetable}{S\arabic{table}}%
        \setcounter{figure}{0}
        \renewcommand{\thefigure}{S\arabic{figure}}%
     }

\newcounter{desccount}

\newcommand{\descref}[1]{\hyperref[#1]{#1}}

\renewcommand{\emptyset}{\varnothing}

\newcommand{\im}{\operatorname{im}}

\theoremstyle{definition}

\makeatletter
\newcommand{\pushright}[1]{\ifmeasuring@#1\else\omit\hfill$\displaystyle#1$\fi\ignorespaces}
\newcommand{\pushleft}[1]{\ifmeasuring@#1\else\omit$\displaystyle#1$\hfill\fi\ignorespaces}
\makeatother

\title{The Emergence of Higher-Order Structure in Scientific and Technological Knowledge  Networks\thanks{E-mail \href{mailto:gebhart@umn.edu}{gebhart@umn.edu} or \href{mailto:rfunk@umn.edu}{rfunk@umn.edu}. We thank the National Science Foundation for financial support of work related to this project (grants 1829168 and 1932596).}}
\author[1]{Thomas Gebhart}
\author[2]{Russell J. Funk}
\affil[1]{Computer Science and Engineering, University of Minnesota}
\affil[2]{Carlson School of Management, University of Minnesota}
\date{}

\begin{document}

\maketitle

\begin{abstract}
The growth of science and technology is a recombinative process, wherein new discoveries and inventions are built from prior knowledge. Yet relatively little is known about the manner in which scientific and technological knowledge develop and coalesce into larger structures that enable or constrain future breakthroughs. Network science has recently emerged as a framework for measuring the structure and dynamics of knowledge. While helpful, existing approaches struggle to capture the global properties of the underlying networks, leading to conflicting observations about the nature of scientific and technological progress. We bridge this methodological gap using tools from algebraic topology to characterize the higher-order structure of knowledge networks in science and technology across scale. We observe rapid growth in the higher-order structure of knowledge in many scientific and technological fields. This growth is not observable using traditional network measures. We further demonstrate that the emergence of higher-order structure coincides with decline in lower-order structure, and has historically far outpaced the corresponding emergence of higher-order structure in scientific and technological collaboration networks. Up to a point, increases in higher-order structure are associated with better outcomes, as measured by the novelty and impact of papers and patents. However, the nature of science and technology produced under higher-order regimes also appears to be qualitatively different from that produced under lower-order ones, with the former exhibiting greater linguistic abstractness and greater tendencies for building upon prior streams of knowledge. 

\end{abstract}

\pagebreak

\section{Introduction}
The past 100 years have witnessed greater progress in science and technology than perhaps any other period in human history. Against this backdrop, observers have expressed concerns over the possibility of science and technology becoming victims of their own success \citep{cowen2019rate, bloom2020ideas}. As science and technology have grown, so has the knowledge researchers must master before arriving at the frontiers of their fields, thereby making future advances slower and more challenging \citep{jones2009burden, milojevic2015quantifying, agrawal2016understanding, pan2018memory, chu2018too}.
A different line of thinking suggests that future developments in science and technology are unlikely to measure up to the past, as many of the most important (and easiest) breakthroughs may have already been made \citep{arbesman2011quantifying, cowen2011great, gordon2017rise}.
While such concerns are not new---Einstein and others, for example, were already remarking on the burden of knowledge in the 1930s---there is growing empirical evidence of several seismic shifts in the social organization of science and technology that align with these views, including the move to team-based production \citep{wuchty2007increasing, wu2019large}, 
greater emphasis on interdisciplinary research \citep{leahey2017prominent}, and the changing structure of careers \citep{jones2010age}.

Recently, studies have devoted increasing attention to developing techniques for characterizing the structure and dynamics of scientific and technological knowledge. While true progress is difficult to measure (and even define), such characterizations are useful both because they enable more systematic ways of documenting change and for the clues they offer into the underlying mechanisms. To date, perhaps the most common approach has been to represent knowledge as a network---where nodes represent concepts, discoveries, or inventions, and edges represent relationships among them---and then to leverage techniques from network science to examine changing patterns of connection over time \citep{uzzi2013atypical, rzhetsky2015choosing, acemoglu2016innovation, christianson2020architecture, dworkin2019emergent}.
Conceptually, this approach is attractive because it maps well onto theories that suggest advances in science and technology result from recombinations of existing knowledge.
\citep{schumpeter1983theory, fleming2001recombinant}
Findings using network approaches support prior observations on the changing structure of scientific and technological knowledge, but they also add important new insight. In addition to growing in volume, scientific and technological knowledge has also become more complex, as measured by the degree of interconnectedness among components \citep{shi2015weaving, varga2019shorter}.
Authors also observe that bridging---a common proxy for innovation, in which an idea spans two or more disconnected areas of a knowledge network---has declined precipitously \citep{mukherjee2016new, foster2015tradition}.

While existing approaches have been helpful, they are limited in several important respects. Critically, they focus on lower-level, dyadic interactions among knowledge components. Given the enormous volumes of prior scientific and technological knowledge, however, discovery today may be more like a game of high-dimensional chess, requiring different lenses for seeing significant moves and comprehending the state of play. Thus, the focus of existing approaches on lower level interactions is one potential explanation for observed slowdowns; progress in science and technology may be playing out in places where current tools are not looking. Indeed, such a possibility could account for observations of slowing progress on macro indicators simultaneously with announcements of major breakthroughs in many fields, including the measurement of gravity waves and advances in deep learning, among others.

In this study, we use methods from algebraic topology to characterize the higher-order structure of knowledge in science and technology, beyond pairwise interactions. Using these methods, we observe rapid growth in the higher-order structure of knowledge in many scientific and technological fields. While dramatic, this transformation is not observable using traditional network measures. The emergence of higher-order structure coincides with a decline in lower-order structure, and has historically outpaced the growth of higher-order structure in scientific and technological collaboration networks. Using persistent homology, we observe increasing topological divergence in the structure of knowledge across fields of science and technology. We find even greater divergence when comparing the structure of knowledge networks to the collaborative networks that produce them, suggesting that the growing complexity of knowledge may be outpacing the collaborative capacities of scientists and technologists. Up to a point, increases in higher-order structure are associated with better scientific and technological outcomes, as measured by novelty and impact. However, as higher-order structure increases, the kind of science and technology produced is qualitatively different from that of lower-dimensional regimes and, by extension, that characteristic of earlier eras. As higher-order structure goes up, papers and patents tend to be both more linguistically abstract and to build more upon existing streams of knowledge. 

Overall, our findings suggest a dual interpretation of the emergence of higher-order structure. The growth in productive scientific and technological knowledge is increasingly taking place in higher dimensions, replacing pairwise relations between granular knowledge areas as drivers of discovery and invention. This greater topological richness may enable the development of entirely new discoveries and inventions, hitherto unknown to science and technology. Yet realizing the opportunities afforded by such higher-order structure and overcoming the challenges it poses is likely to require novel approaches, both for ``doing'' science and technology and measuring their future progress.

\begin{figure}[t]
\includegraphics[width=\textwidth]{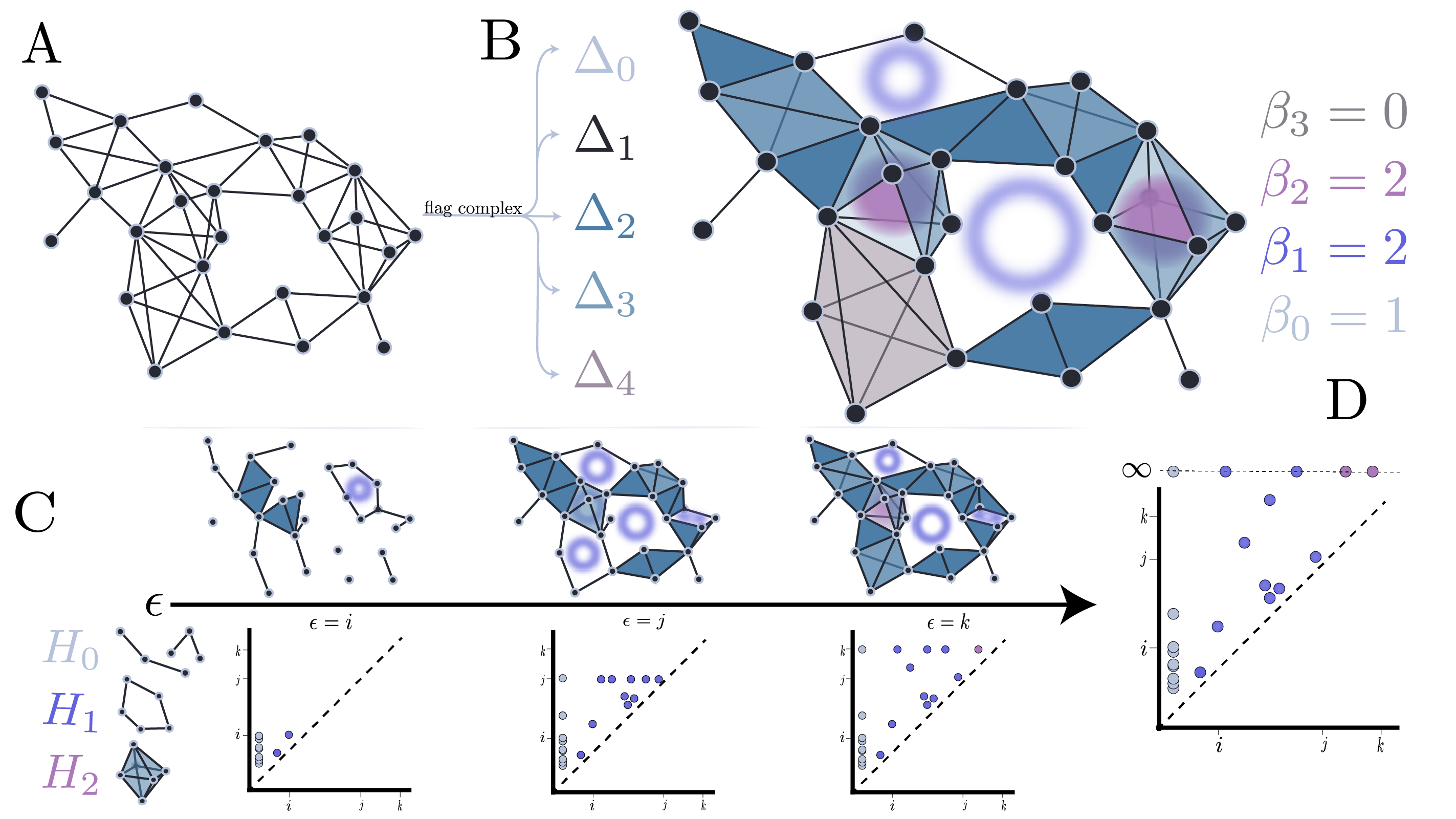}
\caption{\textbf{Persistent homology of complex networks.} \textbf{A}. Graphical representation of a network. \textbf{B}. Transformation of the network into a flag complex. Nodes become 0-cells, edges become 1-cells, 3-cliques become 2-cells, etc. Homological features are denoted by circular and spherical voids. This flag complex contains a single connected component ($\beta_0 = 1$); two 1-dimensional, circular holes ($\beta_1 = 2$); two 2-dimensional, spherical holes ($\beta_2 = 2$); and no higher-dimensional holes ($\beta_3 = 0$). \textbf{C}. Filtration of the complex above varying inverse edge weight threshold $\epsilon$. Topological features in each dimension are tracked on the persistence diagram as $\epsilon$ increases. Point color corresponds to homological dimension. \textbf{D}. After all edges have been added in the filtration, points at infinity correspond to homological features that are never destroyed. These points at infinity correspond to the Betti numbers (topological holes of a given dimension) of the simplicial complex.}
\label{figure:Cartoon}
\end{figure}

\section{Methods}\label{Methods}

We begin this section with a brief overview of homology and persistent homology on networks. For a comprehensive overview of homology and its importance in algebraic topology, see \citet{hatcher2002algebraic}. \citet{carlsson2009topology} and \citet{ghrist2008barcodes} give excellent introductions to applied topology and the usage of persistent homology therein. \citet{aktas2019persistence} reviews the application of these tools to complex networks. We conclude this section with a description of our data our approach to mapping scientific and technological knowledge networks. 

\subsection{Persistent homology}

Assume a weighted graph $G = (V,E,w)$ where $V$ is the set of vertices, $E$ the set of edges, and $w: E \rightarrow \mathbb{R}^+$ a function mapping edges to their weights. Let $w$ define an ordering on the edges of $G$, such that we may decompose $G$ as a sequence of subgraphs $\emptyset = G_0 \subset G_1 \subset \dots \subset G_M = G$ where $M$ is the number of distinct edge weights of $G$. Here $G_1$ is the subgraph of $G$ including only the edge(s) of highest weight. Each inclusion of $G_i$ into $G_{i+1}$ (i.e. $G_i \xhookrightarrow{} G_{i+1}$) is the inclusion of the subgraph $G_i$ into the larger graph $G_{i+1}$ that includes all of $G_i$ and additional edges, with weight equal to the $(i+1)$-th largest weight. At each step $i$ in this decomposition of the graph, we compute the \emph{clique complex} or \emph{flag complex} $K(G_i)$ of $G_i$ by treating each $k$-clique of $G_i$ as a $(k-1)$-dimensional \emph{simplex} or \emph{cell}. In other words, we ``fill in'' all cliques of the graph at each step $i$ such that $0$-cells correspond to nodes, $1$-cells to edges, $2$-cells to $3$-cliques (triangles), $3$-cells to $4$-cliques (tetrahedra), and so on. More generally, each $k$-cell $\sigma = (v_0, v_1, \dots, v_k)$ is the convex hull of $k+1$ affinely-positioned nodes. We denote the set of $k$-cells $\Delta_k$. $K(G_i)$ is an abstract simplicial complex, meaning it is closed under taking subsets of $V$, so any subset of a cell must also be a cell. The filtration of $G$ extends to a filtration of simplicial complexes of $G$ such that $\emptyset = K(G_0) \xhookrightarrow{} K(G_1) \xhookrightarrow{} \dots \xhookrightarrow{} K(G_M) = K(G)$. See Figure \ref{figure:Cartoon} for an example of the construction of simplicial complexes from a graph (B) and the associated filtration (C).  

This filtration and simplicial representation over $G$ is equivalent to the Vietoris-Rips filtration of the point cloud represented by $V$. To see this, consider the pairwise-distance between points $(u,v) \in E$ given by $\frac{1}{w(u,v)}$, such that points not connected in the graph have infinite distance. We can construct the Vietoris-Rips filtration from these distances, and its homology will be identical to the flag complex construction described above \citep{ghrist2008barcodes}.

To track the topological holes of $K(G_i)$, we introduce the chain group, $C_k(K(G_i))$, which is a vector space with basis elements the $k$-cells of $K(G_i)$. Elements of $C_k(K(G_i))$ are linear combinations of these basis elements, and are referred to as $k$-chains. We have freedom in the choice of coefficient field for this vector space, with different choices offering different interpretations of the corresponding homology. We use $\mathbb{Z}_2$ for computational simplicity and ease of interpretation.

To find holes in particular dimensions, we need to know how cells in higher dimensions map onto lower-dimensional cells. We define the \emph{boundary operator} $\partial_{k}: C_{k} \rightarrow C_{k-1}$ as
\begin{equation*}
    \partial_{k}(\sigma) = \sum\limits_{i=0}^k (-1)^i (v_1, v_2, \dots, v_{i-1}, v_{i+1}, \dots, v_k).
\end{equation*}

Note that applying the boundary operator twice yields zero, $\partial_k \circ \partial_{k+1} = 0$. Cycles in dimension $k$ correspond precisely to the elements of $C_k$ that are mapped to zero by $\partial_k$. In other words, the cycles of $C_k$ are elements of $\ker{\partial_k}$. The image of the $(k+1)$-dimensional boundary, $\im \partial_{k+1}$, comprises the $k$-boundaries. Intuitively, $\partial_{k+1}$ takes the interior of a $(k+1)$-dimensional simplex to its boundary. Therefore, $\im \partial_{k+1} \subseteq \ker\partial_k$ and $\partial_k \circ \partial_{k+1} = 0$ as expected. 

Elements of $\ker \partial_k$ that are not in the image of $\partial_{k+1}$ form the $k$-dimensional holes of the simplicial complex. We would like to count the number of $k$-dimensional holes within the simplicial complex, but there may be many cycles which, by the definition of $\partial$, form a boundary of this $k$-dimensional hole. As such, we must associate all cycles enclosing each unique $k$-dimensional hole. In other words, we form an equivalence class of the cycles by associating any two cycles $x,y \in \ker\partial_k$ if $x - y \in \im \partial_{k+1}$. 

We now have the machinery necessary to define the homology groups of simplicial complexes. The homology group in dimension $k$ of simplicial complex $K(G)$ is precisely the group formed by the equivalence classes described above. Formally, $H_k(K(G)) = \ker\partial_k / \im\partial_{k+1}$. The rank of this group $\beta_k = |H_k(K(G))|$ corresponds to the number of $k$-dimensional holes in $K(G)$. We refer to the rank of this group as the $k$-th Betti number. 

Recall that the inclusion relationship induced by edge weightings on the graph, $G_i \subset G_{i+1}$ extends to a filtration on the associated simplicial complexes $K(G_i) \xhookrightarrow{} K(G_{i+1})$. This inclusion relationship over the simplicial structure extends also to the chain groups, such that $C_k(K(G_i)) \xhookrightarrow{} C_k(K(G_{i+1}))$ by mapping basis elements of $C_k(K(G_i))$ to basis elements of $C_k(K(G_{i+1}))$. Note that the homology of $K(G)$ is defined solely in terms of the chain groups. This implies that, for proper choice of basis, we can map cycles to cycles through the induced map $H_k(K(G_i)) \rightarrow H_k(K(G_{i+1}))$. Through this mapping on homology groups, we can track holes across the entire filtration of $G$. We refer to the point in the filration at which a hole appears as its \emph{birth} and the point at which it disappears its \emph{death}. The difference between these quantities \emph{death} - \emph{birth} is called the \emph{lifetime} of the hole. We can encode the global topology of $K(G)$ across its entire filtration conveniently as points in the upper-half plane known as a \emph{persistence diagram} (Figure \ref{figure:Cartoon} C). Each point in the persistence diagram in dimension $k$ corresponds to a $k$-dimensional hole. Points near the diagonal may be considered ``topological noise'' as they represent holes with small lifetimes that are closed soon after they are born. In contrast, points located significantly off the diagonal may be considered meaningful topological features of the space, as they appear early in the filtration and are closed late. Holes that never die over the course of the filtration correspond to points with death time at infinity, and are the elements of $H_k(K(G))$. Their multiplicity is exactly $\beta_k$.

The space of persistence diagrams is endowed with a natural distance metric. Given two persistence diagrams $P_1$ and $P_2$, we define a $p$-norm optimal transport distance as \begin{equation}\label{eq:bottleneck}
    W_p(P_1, P_2) = \inf_{\phi: P_1 \rightarrow P_2} \left(\sum\limits_{x \in P_1} \| x - \phi(x) \|_p \right)^{\frac{1}{p}}.
\end{equation} For $p = \infty$, Equation \ref{eq:bottleneck} is known as the \emph{bottleneck distance} between $P_1$ and $P_2$. The bottleneck distance corresponds to the distance transported between the two farthest points under an optimal transport map. This distance is known to be stable to the underlying topology of the persistence diagrams, such that small changes in topology correspond to small changes in bottleneck distance \citep{cohen2007stability}. Another useful distance between persistence diagrams is the \emph{Wasserstein distance} which corresponds to Equation \ref{eq:bottleneck} with $p=2$. Although it does not enjoy the same stability guarantees, the Wasserstein distance is preferred to the bottleneck distance in our usage due to its increased expressiveness and intuitive $\ell_2$ averaging of the optimal transport map between points in $P_1$ and $P_2$.

\subsection{Data description}
To characterize the higher-order structure of knowledge in science and technology, we collected data from (1) the American Physical Society (``APS data''), consisting of 630,000 scientific articles published between 1893 and 2018, and (2) the U.S. Patent and Trademark Office's (USPTO) Patents View database (``USPTO data''), which covers 6.5 million patents granted from 1976 to 2017. For our purposes, both sources are useful for their knowledge categorization systems. The APS tags manuscripts by subject with between 1 and 5 Physics and Astronomy Classification Scheme (PACS) codes (e.g., 04.30.-w, ``Gravitational waves''; 14.60.Ef, ``Muons''; 05.60.-k, ``Transport processes''). PACS codes are hierarchical, with approximately 7,300 codes at the most granular level. Similar to the APS, the USPTO codes patents by subject using between 1 and 99 classes from its hierarchical, U.S. Patent Classification (USPC) system (e.g., 712/10+, ``Array processors'', 558/486, ``Nitroglycerin''; D13/165, ``Photoelectric cell''). Relative to PACS codes, the USPC system is larger, with more than 158,000 categories at the most granular level. The USPC system is regularly updated to account for changes in technology. With each update, the USPTO reassigns (as necessary) the codes given to all previously granted patents. Thus, the codes assigned to a particular patent may change over time. 

We limit our focus to the period 1980 to 2010 for papers and 1976 to 2010 for patents. PACS codes were introduced in the mid-1970s, but they were not applied consistently in our data for the first few years. In addition, while the USPTO dates to 1790, patent data are only available in machine readable form from 1976. Beginning in the mid-2010s, the APS and USPTO began retiring PACS and USPC codes, respectively, in favor of new systems. We limit our attention to utility patents, which encompass the vast majority (roughly 90\%) of all patents granted by the USPTO. Thus, we exclude from our analysis design patents, plant patents, and reissue patents, which are distinctive in their nature and scope. 

Even after this subsetting, the space of technologies encompassed by the USPTO data is large and heterogeneous, spanning furniture to semiconductors. We therefore characterize knowledge network topology separately for subfields of technology, using National Bureau of Economic Research (NBER) subcategories, of which there are 36 (e.g., ``Biotechnology'', ``Communications''). We occasionally present our results by aggregating to the level of the NBER category, of which there are six (e.g., ``Drugs \& Medical'', ``Computers \& Communications''). Because they are smaller and primarily cover a single discipline (physics), we do not characterize the knowledge network topology separately by field for the APS data. 

To complement our results on knowledge network topology, we also consider collaboration networks in some of our analyses. In collaboration networks, nodes correspond to authors and edges correspond to co-authorship. When mapping instances of authorship to authors, our data pose a challenge because neither patent inventors nor paper authors are assigned unique identifiers at the time of publication, and individuals often list their names inconsistently across their work (e.g., ``Albert Einstein,'' ``A. Einstein''). Moreover, common names (e.g., ``Mary Smith'') may correspond to distinct authors. We therefore map instances of authorship to authors using identifiers assigned via probabilistic name matching algorithms. For patents, we rely on the ``inventor\_ids'' included with the USPTO data. For papers, we implement our own unsupervised algorithm based on a previously described technique \citep{schulz2014exploiting}.

\subsection{Knowledge networks}\label{sec:knowledge_networks}

For each year of the USPTO data, we constructed weighted networks representing the co-occurrence of USPC knowledge classification codes within patents. For APS, we constructed similar weighted networks, with edge weights corresponding to co-occurrence of PACS codes within articles. More formally, let $\mathcal{K}_y^c = (V_y^c,E_y^c,w_{\mathcal{K}})$ represent the network in year $y$ of subcategory $c$ with weight function $w_{\mathcal{K}}$ acting on the edges. Here, $y \in [1975, 2012]$ and possible values of $c$ are listed in Table \ref{table:StatisticsUSPTO_APS}. Letting $V$ represent all knowledge areas across both datasets, the set $V_y^c \subset V$ represents the knowledge areas co-occurring on all works of subcategory $c$ within year $y$. Note that the patent knowledge areas in $V_y^c$ are not distinct across subcategories or years, such that networks of two different subcategories could contain nodes that represent the same knowledge area. 

We define $w_{\mathcal{K}}: E_y^c \rightarrow \mathbb{R}$ as the inverse number of co-occurrences between any two knowledge areas across all works in year $y$ and subcategory $c$. This edge weighting represents the pairwise distances among knowledge areas in the network, so that knowledge areas co-occurring more frequently are closer than knowledge areas that are less frequently co-occurring. With this distance structure, we computed the $k$-dimensional persistent homology for each subcategory network in each year.

\subsection{Collaboration networks}
For the USPTO and APS data, we constructed weighted networks for each subcategory, where weights represent the co-authorship frequency among authors in a particular sliding window. We constructed collaboration networks with both 1 and 3-year sliding windows to align with past work and to match the 1-year sliding window of the knowledge networks. Let $\mathcal{C}_{i,y}^c = (A_{i,y}^c,P_{i,y}^c,w_{\mathcal{C}})$ represent the network in year $y$ with lookback window $i$ of subcategory $c$ with weight function $w_{\mathcal{C}}$ acting on the edges. Here, $y \in [1975, 2012]$, $i \in \{1,3\}$, and possible values of $c$ are listed in Table \ref{table:StatisticsUSPTO_APS}. Letting $A$ be the set of all authors, $A_{i,y}^c \subset A$ represents the authors with works of subcategory $c$ within years $[y-i,y]$. Subsequently, $P_{i,y}^c \subset A_{i,y}^c \times A_{y,i}^c$ represents the relational structure among authors wherein two author nodes are connected by an edge if they were co-authors on a paper in subcategory $c$ in years $[y-i,y]$. Note that the authors $A_y^c$ are not distinct across subcategories or years. 

We define $w_{\mathcal{C}} : P_{i,y}^c \rightarrow \mathbb{R}$ as the inverse of the number of co-authored papers between any two authors for a given $c, y$, and $i$. This edge weighting represents the pairwise distances among authors in $\mathcal{C}_{i,y}^c$, such that more frequent co-authors are ``closer'' than less frequent co-authors. With this distance structure, we computed the $k$-dimensional persistent homology for each subcategory network in each year.

\section{Interpreting knowledge network topology}\label{sec:interpreting_topology}

In converting a network to its flag complex, whenever a $k$-clique is formed, we ``fill it in'' and treat the clique as a higher-order topological object: a $(k-1)$-dimensional cell. We may view these higher-order cells as abstract knowledge areas composed of $k$ granular knowledge areas. Higher-dimensional homology tracks how these abstract knowledge areas combine, qualifying the meso- and macro-scale structure in the knowledge networks by enumerating the higher-dimensional homology groups which correspond to ``holes.'' In the first dimension, $\beta_0$ corresponds precisely to the number of connected components of the network. Higher-dimensional holes $\beta_{i > 0}$ imply a relative lack of connectivity across the abstract knowledge areas determined by the $(i+1)$-dimensional cells, just as $\beta_0$ implied a lack of connectivity between clusters of granular knowledge areas. Homology provides a global characterization of structure across the hierarchical representation of knowledge provided by the underlying flag complex of the knowledge network.

Persistent homology tracks this knowledge structure through a filtration on the edge weights of the network, and provides a more granular characterization of this structure across scales of co-occurrence frequency. The relational structure of the knowledge networks is determined by the frequency of co-occurrence of knowledge areas within works. Using inverse frequency as the filtration parameter, we can view the flag filtration as a discrete assembly of the network, wherein the edge connecting the most commonly co-cited knowledge areas is added to the network first, followed by the second most commonly co-cited pair, and so on. This construction process continues until the last step in the filtration, where edges between knowledge areas with the fewest co-occurrences are added. The interpretation of highly-weighted $1$-cells (edges) as pairs of knowledge areas that are frequently combined within works extends to higher dimensions. For example, a $2$-cell having high weight on all edges (thus appearing early in the filtration) represents three knowledge areas that are frequently combined within a subfield. Decomposing the flag complex across co-occurrence frequency provides further insight into the meso- and macro-scale structure of these networks apart from their Betti numbers. For example, homology by itself cannot distinguish between a network that retains a large number of connected components up until a small threshold value of co-occurrence frequency in which they all join and a network that grows as a single, dense component across all threshold values. Persistent homology can differentiate these networks, and this difference would be evident in their persistence diagrams.

\section{Results}

We begin by documenting the emergence of higher-order structure across fields. We show that this emergence is not observable using traditional network measures. Subsequently, we relate the topological structure of fields to the nature of the science and technology they produce. We find a striking relationship between higher-order structure and the linguistic abstractness of words used within fields over time. Finally, we observe robust relationships between the higher-order structure of knowledge and measures of to discovery and invention in papers and patents.

\subsection{Emergence of higher-order structure}

\begin{figure}[t]
\includegraphics[width=\textwidth]{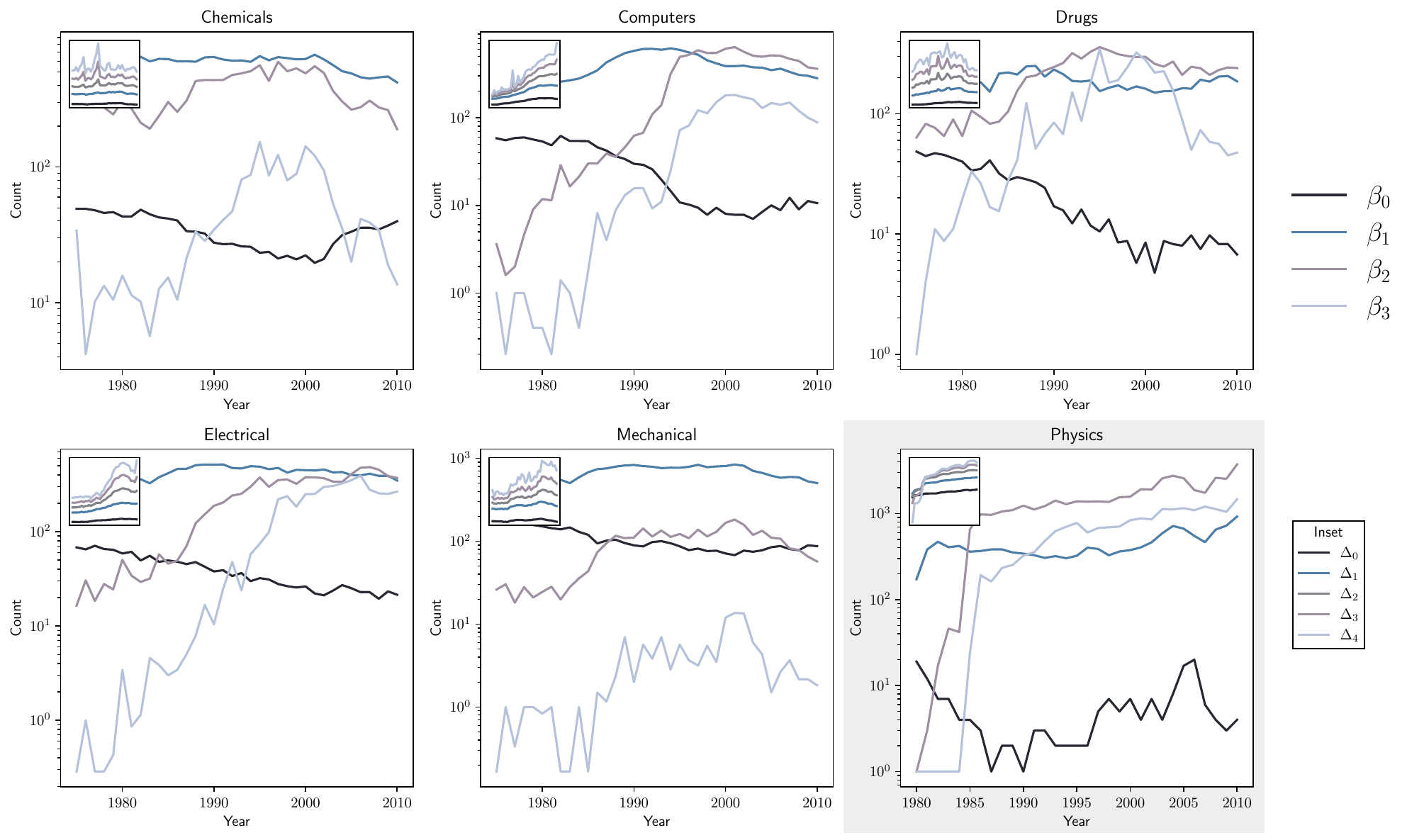}
\caption{\textbf{Knowledge network topology over time at the field level.} The main plots track counts of Betti numbers; inset plots track cell counts. All y-axes are reported on a log scale. The ``Physics'' panel is highlighted to indicate the different data source (APS) and publication type (academic papers) relative to those of the other panels (where data come from the USPTO and the publication type is patents). Note that the underlying topological features are measured at the subfield level; to generate these field-level plots, we report the average values observed for the constituent subfields.}
\label{figure:BettiLog10CategoryDynamics}
\end{figure}

\begin{figure}[htbp]
\includegraphics[width=\textwidth]{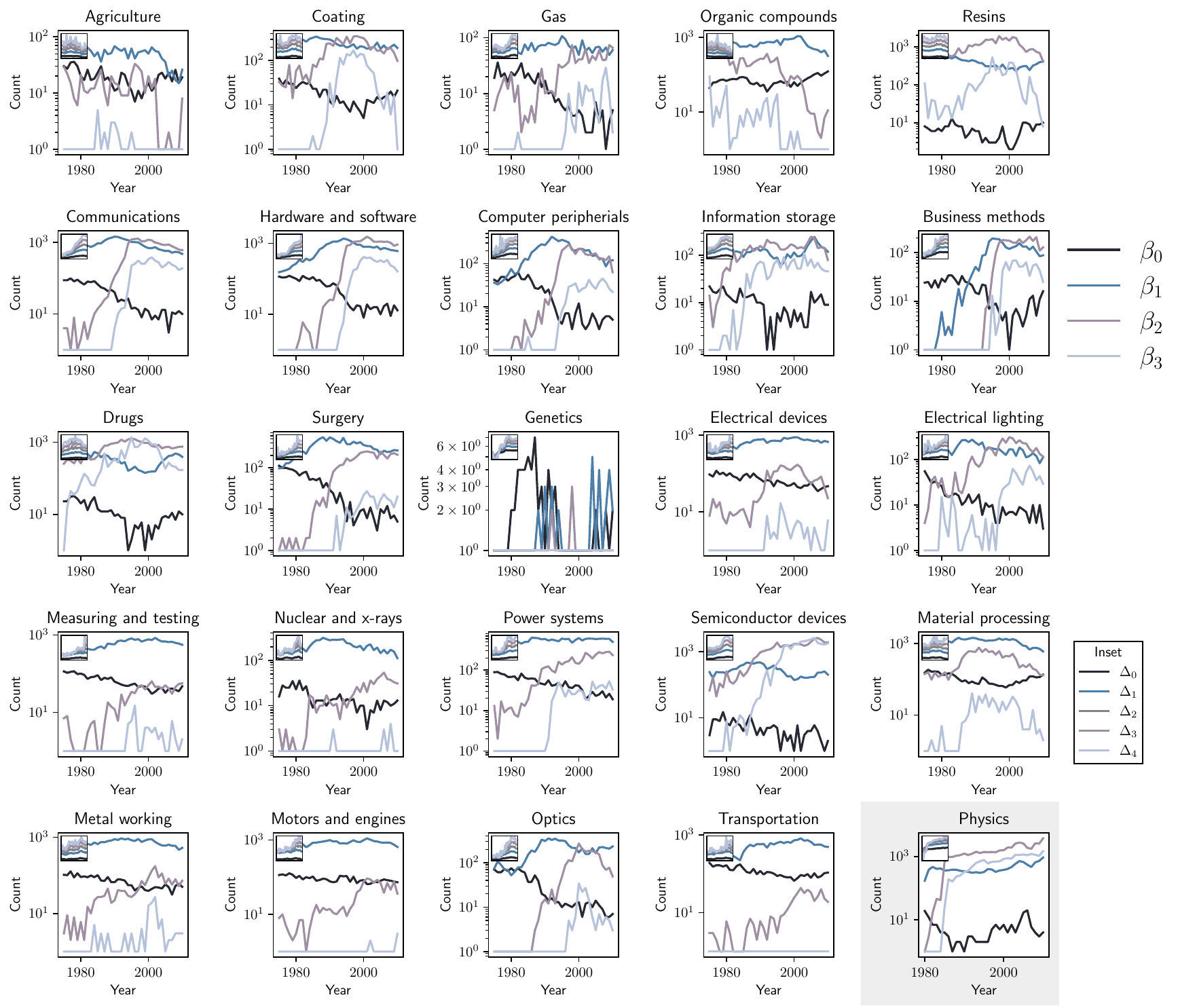}
\caption{\textbf{Knowledge network topology over time at the subfield level}. The main plots track counts of Betti numbers, while the inset plots track cell counts. All y-axes are reported on a log scale. The ``Physics'' panel is highlighted to indicate the different data source (APS) and publication type (academic papers) relative to those of the other panels (where data come from the USPTO and the publication type is patents.}
\label{figure:BettiLog10SubcategoryDynamics}
\end{figure}

We observe substantial shifts in the structure of scientific and technological knowledge. Figure \ref{figure:BettiLog10CategoryDynamics} gives an overview of these changes by plotting the average number of topological holes by year for six major fields of science and technology (Chemicals, Computers, Drugs, Electrical, Mechanical, and Physics). Averages are based on the number of topological holes observed for the constituent subfields. Raw numbers of holes for each subfield are shown in Figure \ref{figure:BettiLog10SubcategoryDynamics}. Corresponding plots for cell counts are shown in the inset axes of Figures \ref{figure:BettiLog10CategoryDynamics} and \ref{figure:BettiLog10SubcategoryDynamics}. The y-axes of all plots are reported on a $\log_{10}$ scale. Darker lines in each plot correspond to lower-order structure; higher-order structure is indicated by lighter colored lines. For display purposes, we exclude fields and subfields categorized as ``other'' from our figures, although they are included in our statistical analyses. 

Beginning with Figure \ref{figure:BettiLog10CategoryDynamics}, we observe several noteworthy patterns. Across fields, there is persistent decline in $\beta_0$, followed by a leveling off (Computers, Drugs, Electrical, Mechanical) or modest increase (Chemicals, Physics) in more recent years (beginning in the early 2000s). Because $\beta_0$ captures the number of connected components in a network, this pattern suggests that with respect to lower-level structure, the knowledge networks of science and technology are becoming more connected over time, a finding that is consistent with observations from prior research \citep{varga2019shorter}.

Turning to $\beta_1$, we also observe consistent patterns across fields. Relative to $\beta_0$, there is much less change over time; we see modest increases in $\beta_1$ in the  Chemicals, Computers, Drugs, Electrical, and Mechanical fields, typically peaking sometime between the mid-1980s and mid-1990s, before declining to levels similar to those observed in the 1970s. Physics is an exception to this pattern; there, we observe a dramatic increase in $\beta_1$ beginning in the late 1970s, which ends in a local peak in the early 1980s, after which there is a gradual dip, followed by a similarly gradual increase. Given the growth of $\Delta_2$ (triangles) over time, all fields but Chemistry and Drugs have seen increases in the number of abstract knowledge areas composed by triplets of low-level concepts. The relatively constant value of $\beta_1$ over time implies that either $1$-dimensional holes are being created about as often as they are being closed, or that these knowledge areas are being generated from pre-existing knowledge structures, such that few holes are created or closed. 

Relative to $\beta_0$ and $\beta_1$, we see much more variation, both within and between fields, for higher dimensional holes (i.e., $\beta_2$ and $\beta_3$). Beginning with $\beta_2$, the Chemicals, Drugs, and Mechanical fields follow similar trajectories, with moderately large numbers of 2-dimensional holes (compared to other fields), followed by a gradual increase, and then a leveling off or slight decrease. The pattern of change is more dramatic for the Computers, Electrical, and Physics fields, where we observe relatively low counts of $\beta_2$ in the early years, after which are dramatic increases, followed by a leveling off (but no decline). Finally, with respect to 3-dimensional holes, the patterns are generally similar to those of $\beta_2$,  with the exception of Drugs, which has a much more dramatic increase, akin to the fields of Computers, Electrical, and Physics. The dramatic increase in $\beta_2$ and $\beta_3$ in tandem with an increase in $\Delta_3$ and $\Delta_4$ implies that high-level knowledge areas are being created but are only recently becoming synthesized. This is in contrast to the Mechanical field, wherein abstract knowledge continues to be constructed (given the increasing cell counts in high dimensions) but without introducing new gaps as implied by the steady $\beta_2$ and $\beta_3$.

Increases in higher-order structure often coincide with decreases in lower order structure. This pattern is evident in the crossovers between lines tracking holes of different dimensions. Consider Computers, where the relative ranking of holes by commonality changes several times over the study period. In the 1970s, the most prevalent holes are 1-dimensional, followed by dimensions 0, 2, and 3. By the late 1990s, the ordering has shifted, with the most prevalent holes now being those of dimension 2, followed by dimensions 1, 3, and 0. This inverse relationship between higher- and lower-order structure is important because it suggests not only that discovery may be increasingly playing out at higher levels, but also that it is doing so less at lower levels, and may therefore be less visible using traditional techniques. Overall then, these patterns are consistent with our claims on the growing importance of higher-order structure.

Turning to Figure \ref{figure:BettiLog10SubcategoryDynamics}, we observe similar patterns at the subfield level. The shift from lower- to higher-order structure is visible across many different and diverse subfields, as evidenced by dramatic crossovers in lines tracking different dimensions of holes. There are also, however, several interesting departures from the general patterns. For example, while by the end of the study window, many subfields have noteworthy counts of 3-dimensional holes, there is substantial variation in when those holes appear. Organic Compounds seems to buck the general transition to higher-order structure; over the study period, with lower-dimensional holes generally becoming more prevalent.

Notwithstanding their mathematical connection, patterns of change in cell counts (inset axes) are distinctive from those of topological holes. The relative distribution of cell counts is fairly stable; we see fewer crossovers, and the ranks of cells by commonality tends to correspond to dimension (with higher dimensional cells being most common). The overall trajectories of the Computers, Electrical, Mechanical, and Physics fields are fairly similar, being stable in earlier years, followed by dramatic growth and subsequent leveling off (Computers, Physics) or decline (Electrical, Mechanical).

The emergence of higher-order structure often coincides with historical events in the underlying fields. Consider the Computers and Electrical categories, which experienced dramatic growth in higher-order structure from the early-1980s through the mid-1990s. For both fields, the period was one of dramatic technological change, witnessing the invention of flash memory (1980), the scanning tunneling microscope (1981), World Wide Web (1990), carbon nanotubes (1991), and the Mosaic web browser (1993). Drugs also offers a telling illustration. There, increases in higher-order structure (particularly 3-dimensional holes) map closely to major breakthroughs in biotechnology, including marketing of the first recombinant DNA drug (Humulin, 1983) and the invention of the polymerase chain reaction (1983).

\begin{figure}[!htbp]
\includegraphics[width=\textwidth]{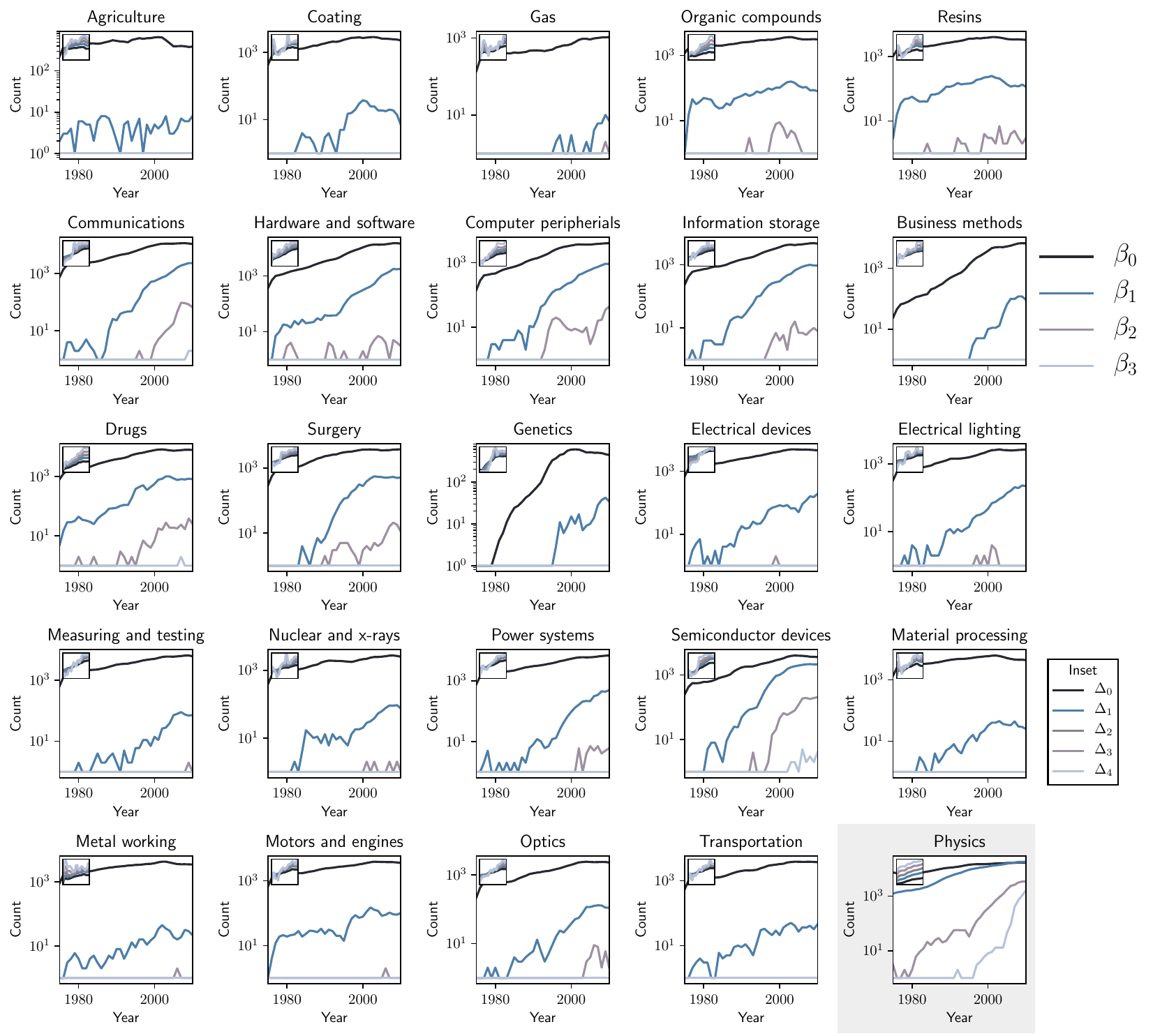}
\caption{\textbf{Plots of collaboration network topology over time (3 year moving window) Subfield level.} The main plots track counts of Betti numbers, while the inset plots track cell counts. All y-axes are reported on a log scale. The ``Physics'' panel is highlighted to indicate the different data source (APS) and publication type (academic papers) relative to those of the other panels (where data come from the USPTO and the publication type is patents). In contrast to approach for knowledge networks (where we use a 1 year window), we define collaboration networks using a 3-year moving window.}
\label{figure:BettiLog10SubcategoryDynamicsCollab3yr}
\end{figure}

The sharp increase in higher-order structure is interesting when contrasted with the more tempered growth in the corresponding collaboration networks over the same period. Figure \ref{figure:BettiLog10CategoryDynamicsCollab3yr} shows that across fields, scientists and technologists are increasingly forming higher-order collections of collaborative groups, but these changing collaboration patterns do not map closely to changes in the topology of knowledge. At the subcategory level, Figure \ref{figure:BettiLog10SubcategoryDynamicsCollab3yr} reinforces this observation, showing that many fields have only recently seen increases  in $\beta_2$ and $\beta_3$. The number of connected groups of researchers ($\beta_0$) appears to be stable or increasing slightly over time. This is in contrast to the generally declining number of connected components observed in the knowledge networks (Figure \ref{figure:BettiLog10CategoryDynamics}). 

To ensure the above historical trends are not a feature of the knowledge network construction itself or whether all two-mode networks of this size show these dynamics, we repeated these analyses on a completely different source of knowledge networks constructed via question category tags across a number of Stack Exchange sites. Notably, we observe a slight upward trend in higher-dimensional holes over time, but this growth is much slower than that of the science and technology networks presented above. See the supplementary Section \ref{sec:additional_knowledge_networks} for more information. We also compared the homological distributions the knowledge networks to those of three random graph models of equivalent size across a range of parameters (Section 
\ref{sec:random_comparison}). The results of this analysis, displayed in Figure \ref{figure:RandomNetworkKnowledgeMatch}, indicate that the structure of the knowledge networks studied in this work are non-random and are not easily described by simple construction rules like small-worldness \citep{watts1998collective} or preferential attachment \citep{barabasi1999emergence}.

\subsection{Diverging topologies}

\begin{figure}[t]
\includegraphics[width=\textwidth]{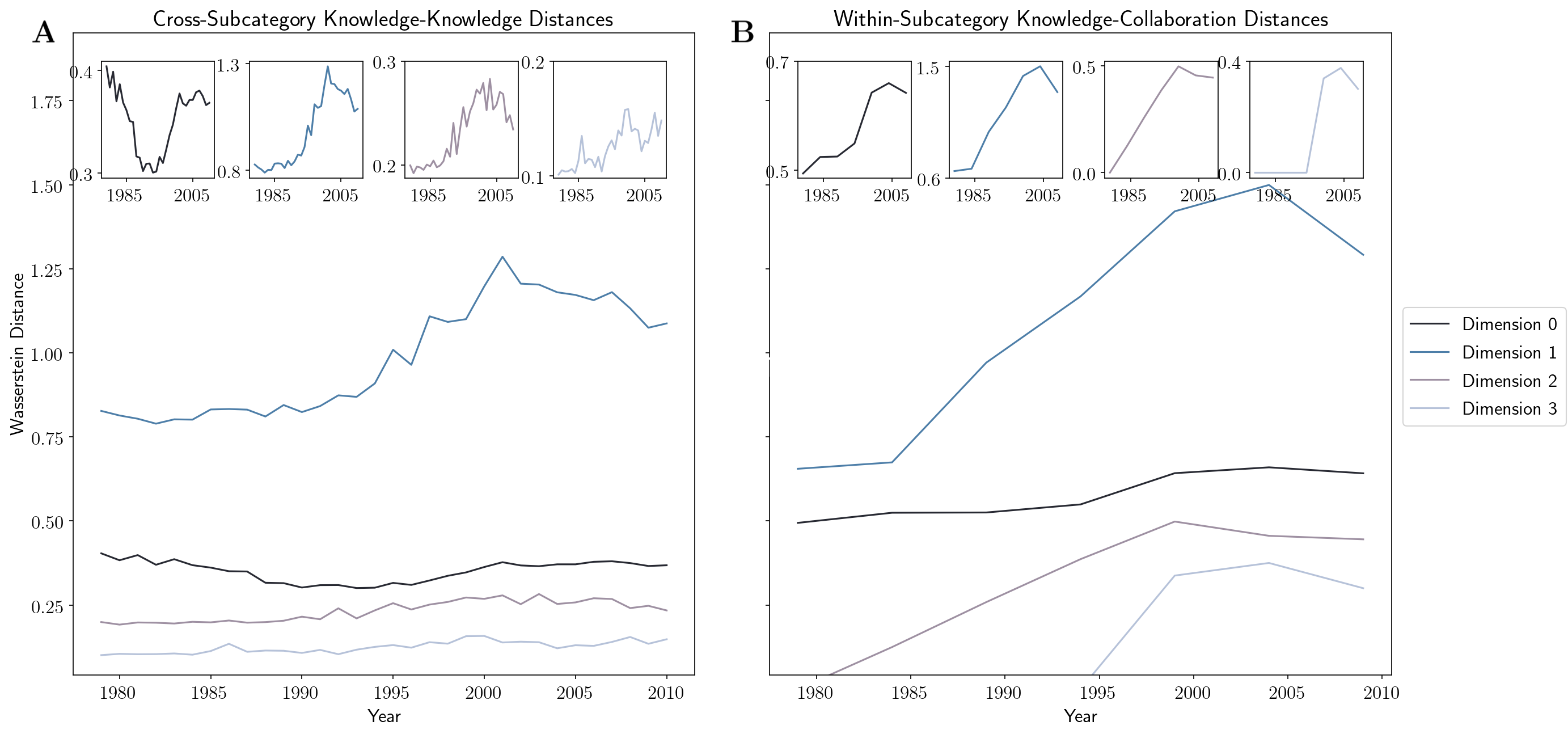}
\caption{\textbf{Cross- and within-subcategory diagram distances.} \textbf{A}: Mean Wasserstein distances between persistence diagrams of knowledge areas in each year. For each year and dimension, the Wasserstein distance was calculated between each knowledge network across each subcategory. The mean of these distances in each year and dimension shown. Inset plots show individual dimensions. \textbf{B}: Mean Wasserstein distances between persistence diagrams of knowledge networks and collaboration networks of a given subcategory. For each year and dimension, the Wasserstein distance was calculated between each knowledge network and its corresponding collaboration network within a particular subcategory. The mean of these distances in each year and dimension is shown. Inset plots show individual dimensions.}
\label{figure:KnowlKnowlCollabDistances}
\end{figure}

Given this overall increase in higher-order structure, we would like to know whether this reflects a topological change over time that is structurally similar across all fields or whether these structural changes are field-specific. To do this, we compute the Wasserstein distance (Equation \ref{eq:bottleneck}) between persistence diagrams in a pairwise fashion between each knowledge subcategory in each year. While individual knowledge areas may converge in their topological similarity across a number of years, the aggregate difference in topological structure between knowledge areas points to significant divergence over time.  Panel A in Figure \ref{figure:KnowlKnowlCollabDistances} depicts the average Wasserstein distance across all subcategory pairs in a given year with inset plots showing individual dimensions on their own scale. Knowledge subcategories differ most in their 1-dimensional persistent homology, with growing divergence until the mid-2000's. In more recent years, topological divergence has increased in higher dimensions. These results, combined with the homological trends over time, imply that the dimensionality of scientific and technological knowledge is both increasing and becoming more topologically heterogeneous across fields.

\begin{figure}[htbp]
\includegraphics[width=\textwidth]{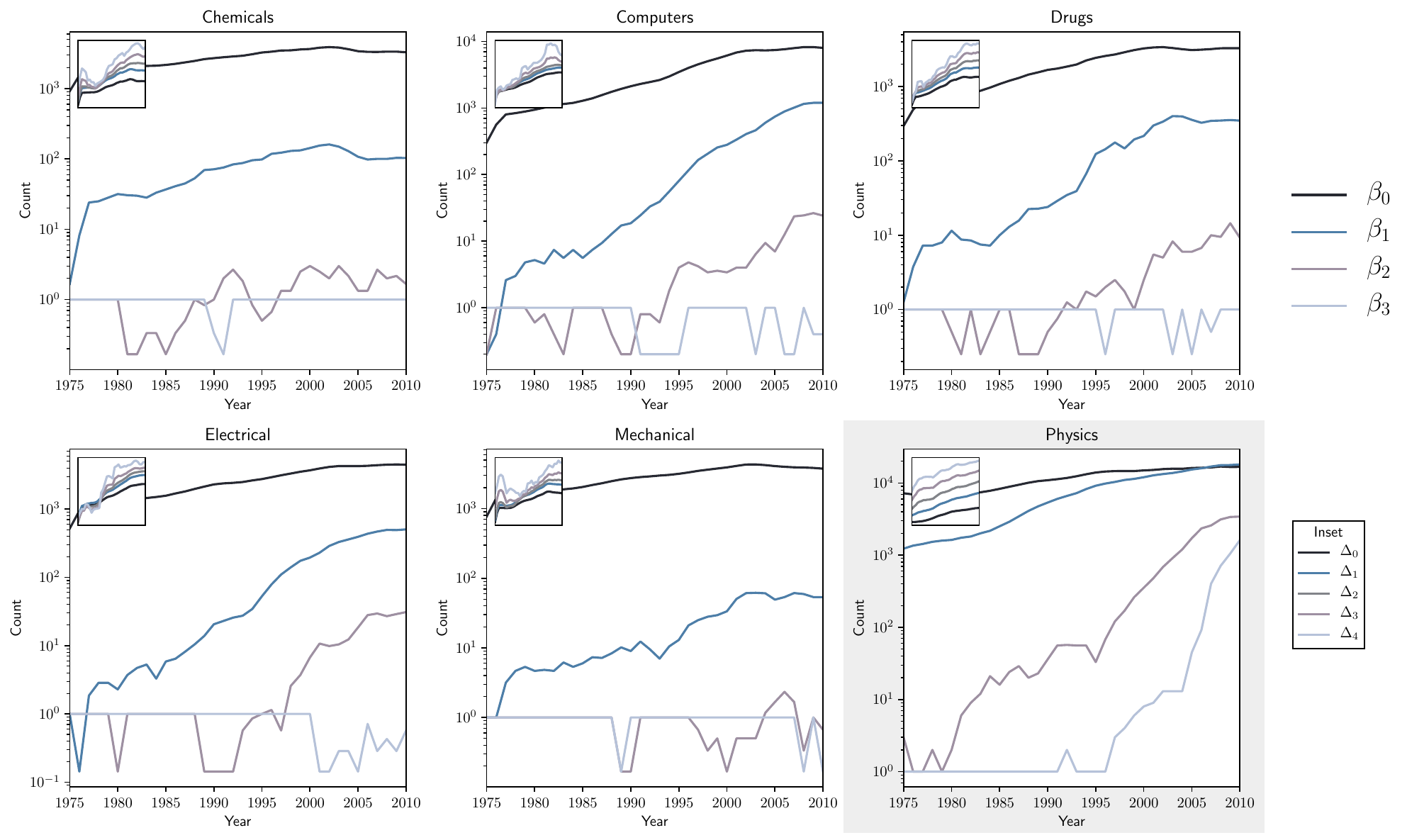}
\caption{\textbf{Collaboration network topology over time at the field level.} The main plots track counts of Betti numbers, while the inset plots track cell counts. All y-axes are reported on a log scale. The ``Physics'' panel is highlighted to indicate the different data source (APS) and publication type (academic papers) relative to those of the other panels (where data come from the USPTO and the publication type is patents). Note that underlying topological features are measured at the subfield level; to generate these field-level plots, we report the average values observed for the constituent subfields. In contrast to approach for knowledge networks (where we use a 1 year window), we define collaboration networks using a 3-year moving window.}
\label{figure:BettiLog10CategoryDynamicsCollab3yr}
\end{figure}

We also examine whether the emergence of higher-order structure is visible in the collaboration networks of scientists and technologists. Panel B of Figure \ref{figure:KnowlKnowlCollabDistances} plots the average distance between the knowledge and (3-year) collaboration network persistence diagrams, within each subcategory, across time. Individual dimensions are again plotted in the inset. The topological structures of knowledge and collaboration networks have diverged dramatically over time, leveling out only recently. This divergence appears to result from the emergence of higher-order structure in the collaboration networks of scientists and technologists severely lagging that of the corresponding knowledge networks, with higher-dimensional holes beginning to appear in the latter only in the past decade (Figure \ref{figure:BettiLog10CategoryDynamicsCollab3yr}). This recent increase in higher-order collaborative structure may be driven at least in part by the growing complexity of knowledge, as progress on open problems more and more requires attention from larger teams of researchers with expertise spanning multiple domains \citep{wuchty2007increasing, porter2009science, varga2019shorter}.

\subsection{Associations with measures of lower-order network structure}

\begin{figure}[t]
\centering
\includegraphics[width=5in]{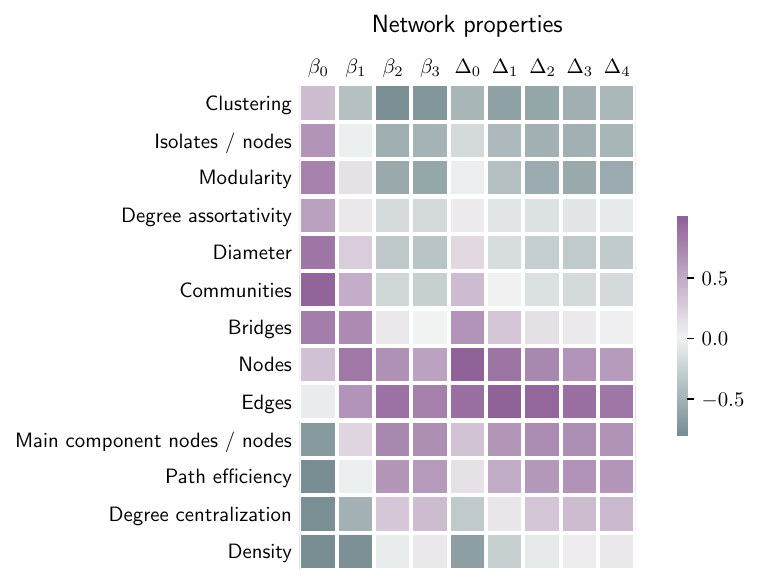}
\caption{\textbf{Correlations between knowledge network topology and some popular network-theoretical measures.} All correlations are statistically significant at the $p<0.05$ level with the exception of isolates / nodes ($\beta_1$), modularity ($\Delta_0$), degree assortativity ($\Delta_0$), communities ($\Delta_1$), bridges ($\beta_3$, $\Delta_4$), edges ($\beta_0$), path efficiency ($beta_1$), and density ($\beta_2$, $\Delta_3$). As expected, network measures that approximate global connectivity (e.g., path efficiency, degree centralization) show higher correlation with higher-dimensional Betti numbers, but this correlation is modest and inverts in lower dimensions. Many of the traditional network measures are uncorrelated with higher-dimensional homology.}
\label{figure:NetworkCorrelations}
\end{figure}

Figure \ref{figure:NetworkCorrelations} visualizes correlations between our measures of knowledge network topology (rows) and several common (lower-order) measures of network structure (columns). The order of the rows was determined using a hierarchical clustering algorithm, such that measures that show similar patterns of correlation with topological properties appear adjacent to one another. As may be expected, counts of nodes and edges are among those most strongly associated with knowledge network topology. Most other measures show relatively low correlation. Notably, correlations with network density are relatively low, which suggests that the emergence of higher-order structure is unlikely an artifact of changes in overall connectivity. We also observe that correlations with clustering are relatively low, which is also reassuring given the conceptual similarities between triadic closure and (lower order) measures of network topology. Finally, observe that the correlations between knowledge network topology and measures of community structure are relatively low. This pattern is noteworthy because community detection is arguably one of the few widely used measures that captures some dimension of higher-order structure. 

\begin{figure}[htbp]
\includegraphics[width=\textwidth]{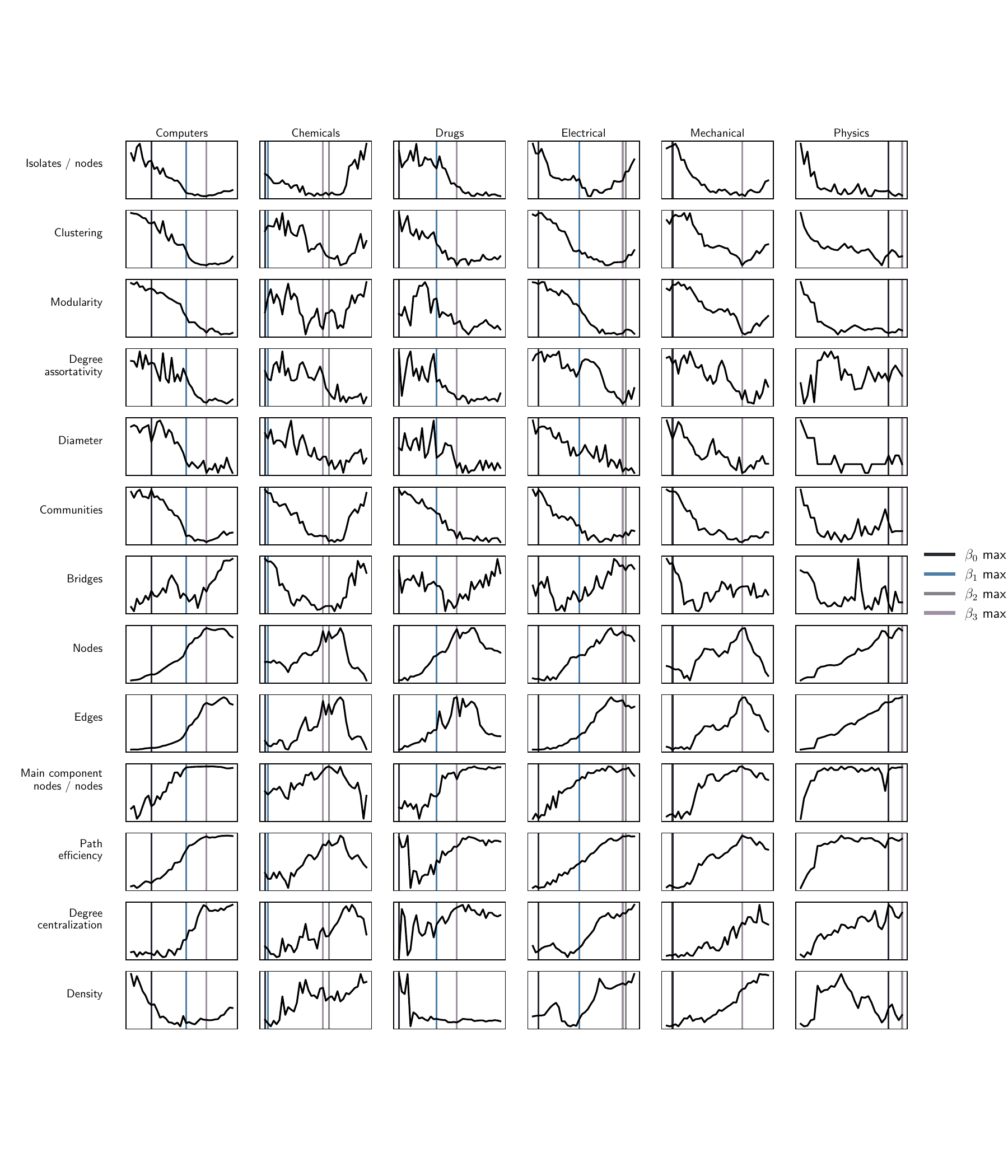}
\caption{\textbf{The evolution of higher-order topology is distinct from traditional network-theoretic measures across time}. Each row depicts a traditional network-theoretic measure popular in prior works. Columns correspond to fields. Maximum Betti number in each dimension depicted as vertical lines. The homological information is distinct from the network-theoretic measures and are not described by these lower-level network properties.}
\label{figure:NetworkCorrelatio }
\label{figure:traditional_networks_over_time}
\end{figure}

Figure \ref{figure:traditional_networks_over_time} plots the measures of network structure from Figure \ref{figure:NetworkCorrelations} over time. Compared to Figure \ref{figure:BettiLog10CategoryDynamics}, we observe that traditional network measures show distinct patterns of temporal evolution. Across fields, there is relatively little consistent alignment between common measures of network structure (e.g., clustering, degree assortativity, and density) and our higher-order measures. We also see relatively little consistent correspondence between our knowledge network topology and network measures that include some higher-order information (e.g., degree centralization, bridges, modularity). Consistent with our interpretation of the correlational data in Figure \ref{figure:NetworkCorrelations}, these results suggest that knowledge network topology encodes information that is not captured in more traditional (lower-order) network measures. 

\subsection{Linguistic properties of science and technology}

To gain additional insight on the meaning of higher-order structure, we conducted an analysis in which we evaluated for differences in word usage by publications as a function of the distribution of Betti numbers for their fields and years of publication. We tokenized words within abstracts of the USPTO dataset, classifying them by their parts of speech. See the Supplemental Materials for more information. 

For each part of speech $\times$ token, we computed the Spearman rank correlation between the number of patents using each part of speech $\times$ token, by subfield $\times$ year observations, separately for the four Betti dimensions. Table \ref{table:AbstractWords} reports a subset of the results of this analysis, showing the top 10 lemmas with the largest (positive) and statistically significant (p<0.05) Spearman rank correlation by part of speech and Betti dimension. To minimize noise, we limit our reporting to lemmas that appeared in the abstracts of at least at least 1000 patents across the entire sample. We found similar results (not reported, but available upon request) using more complex models, including those with adjustment for field and year. We also replicated our analysis using patent titles (rather than abstracts) and found similar patterns. 

The results show noticeable differences in word usage across dimensions. Across all four parts of speech, lower dimensional holes tend to be associated with more concrete lemmas, while higher dimensional holes tend to be associated with more abstract ones. This pattern is particularly clear for nouns. Under $\beta_0$, the most predictive lemmas refer to things that are readily perceptible through sight, sound, and touch (e.g., ``crank’’, ``spool’’, ``shoe’’); strikingly, the list includes 3 of the 6 simple machines (``lever’’, ``wheel’’, ``pulley’’). Under $\beta_3$, by contrast, the most strongly associated nouns refer to things that are less perceptible  (e.g., ``process,'' ``property,'' ``method''). The most predictive nouns for $\beta_1$ and $\beta_2$ encompass a mix of concrete and abstract things. Results for verbs, adjectives, and adverbs follow a similar pattern to nouns. Lower dimensional holes are associated with more concrete lemmas, often indicating direction or mechanical motion (e.g., verbs: ``swing’’, ``pivot’’, ``disengage’’; adjectives: ``slidable’’, ``moveable’’, ``swinging’’; adverbs: ``forwardly’’, ``upwardly’’, ``rotatably’’), while higher dimensional holes are associated with more abstract ones that are typically less amenable to sensory perception (e.g., verbs: ``base’’, ``describe’’, ``contain’’; adjectives: ``present’’, ``active’’, ``more’’; adverbs: ``highly’’, ``significantly’’, ``specifically’’).

\begin{figure}[htbp]
\includegraphics[width=5in]{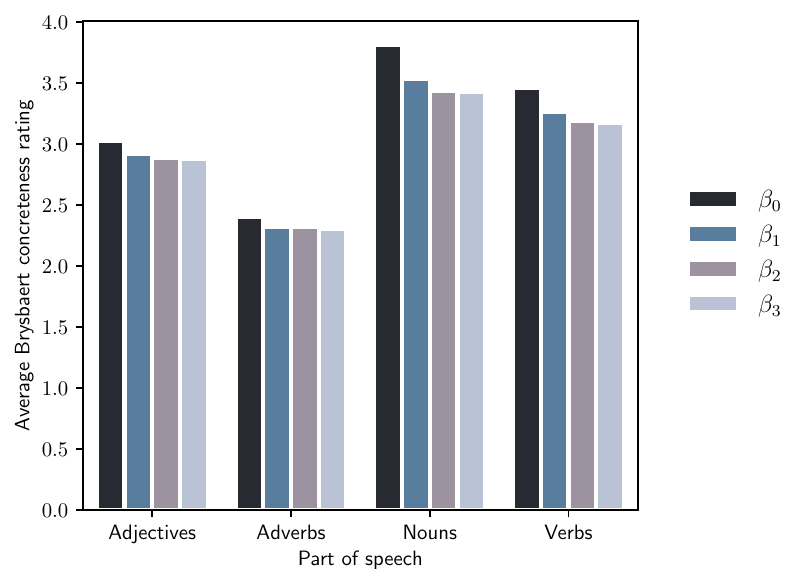}
\caption{\textbf{Distribution of average Brysbaert lexical concreteness across homological dimension and part of speech.} Concreteness of words used in abstracts within a field falls as high-dimensional Betti numbers increase.}
\label{figure:LexicalConcreteness}
\end{figure}

In supplemental analyses, we explored differences in the use abstract terms across Betti dimensions more systematically by assigning each lemma a concreteness score, using the ratings of \citet{brysbaert2014concreteness}. We then pulled lemmas that were significantly (p<0.05) and positively associated with each Betti dimension and computed their average concreteness by dimension and part of speech (Figure \ref{figure:LexicalConcreteness}). Across all four parts of speech, mean concreteness declines with increases in dimension.

We observe that many of the lemmas most associated with higher-dimensional holes are indicative of engagement with (particularly improvements on) prior knowledge. This pattern is especially clear for verbs, where the lemmas most strongly associated with $\beta_3$ include ``base’’, ``enhance’’, ``improve’’, ``add’’, ``use’’, and ``modify’’. We also see evidence of this pattern with adverbs, where ``highly’’, ``efficiently’’, ``significantly’’, and ``optionally’’ are among the lemmas most strongly associated with $\beta_3$.

This increase in abstractness of language with increasing high-dimensional structure aligns with our intuition regarding the interpretation high-order structure within knowledge networks (Section \ref{sec:interpreting_topology}). High-dimensional holes are created through the adjunction of high-dimensional cells, which themselves are composed of a number of granular knowledge areas represented by nodes, and growth in the creation and combination of high-order cells gives rise to high-dimensional structure as measured by the Betti numbers of the network. The results described in this section imply that as scientists and technologists combine granular knowledge areas into abstract knowledge at the level of cells, the language they use mimics this move to abstraction, as they grapple with the complexities implicit in leveraging multiple distinct knowledge areas within a single work. 

\subsection{Models of discovery and invention}

\begin{figure}[t]
\includegraphics[width=\textwidth]{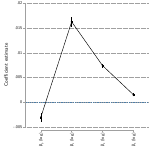}
\caption{\textbf{Plots of coefficients from regressions predicting ``hit'' publications \\ Betti numbers.}  This figure visualizes coefficient estimates for Betti numbers from Table~\ref{table:MainRegressions} (Model 5). Error bars represent 95\% confidence intervals. Error bars that span 0 indicate that the corresponding coefficient is not significantly different from 0; coefficients with overlapping error bars may however be significantly different from each other.}
\label{figure:MainRegressionsCoefficientPlotHoles}
\end{figure}

While our results show dramatic growth in the higher-order structure of scientific and technological knowledge, the implications of this growth are less clear. On the one hand, larger numbers of higher dimensional holes may indicate that future advances will hinge on addressing increasingly difficult problems. On the other hand, higher-order structure may create opportunities for new kinds of breakthroughs, allowing investigators to see and do things not possible within the confines of lower-dimensional knowledge networks.   

To evaluate these possibilities, we estimated regression models that predicted the probability of a publication being in the top 5\% of the citation distribution as a function of the topological properties of its field. Our presumption is that high citation counts are indicative of breakthroughs; we recognize however that citations are an imperfect proxy, and below we demonstrate consistent results using alternative measures.

For patents, we measure topological properties based on the year of application, which typically corresponds to the time of invention; for papers, we measure based on the year of publication. The primary predictors of interest---measured at the field $\times$ year level---are counts of holes by dimension (i.e., $\beta_0$, $\beta_1$, $\beta_2$, $\beta_3$), which we log transform to account for diminishing effects of large counts. We also included counts of cells by dimension, (i.e., $\Delta_0$, $\Delta_1$, $\Delta_2$, $\Delta_3$, $\Delta_4$; log transformed), standard measures of network structure (i.e., density, clustering, and path efficiency), and a measure of publication volume, again all measured at the field $\times$ year level. To account for additional confounding, our models include fixed effects for field and year. All tests of significance are based on heteroskedasticity-robust standard errors. Table \ref{table:SummaryStatistics} shows descriptive statistics for the variables in the models.

Coefficient estimates are shown in Table \ref{table:MainRegressions}. To facilitate interpretation, Figure \ref{figure:MainRegressionsCoefficientPlotHoles} plots these estimates for $\beta_0$ through $\beta_3$ based on Table \ref{table:MainRegressions}, Model 5. We observe a curvilinear (inverted-U shaped) relationship between knowledge network topology and the probability of a hit publication. Increases in 0-dimensional holes ($\beta_0$) are negatively associated with the probability of a hit publication ($\textrm{coef.} = -0.0031; P<0.001$). Higher dimensional holes ($\beta_1$, $\beta_2$, $\beta_3$) are all positively (and significantly) associated with the probability of a hit, with the magnitude of the coefficient being largest for $\beta_1$ ($\textrm{coef.} = -0.0163; P<0.001$) and declining thereafter. Overall, these results offer some support for both views discussed above; increases in higher-order structure are associated with breakthroughs, but primarily at moderate dimensionality.

We conducted several analyses to evaluate the predictive power of topological properties relative to alternative measures. First, the bottom of Table \ref{table:SummaryStatistics} reports Wald tests that evaluate whether the inclusion of topological measures improves model fit. Across all models, including those with field and year fixed effects and control variables, the null hypothesis is rejected; fit improves significantly with the inclusion of the topological measures. Second, we decomposed the adjusted-$R^2$ of Model 5 of Table \ref{table:SummaryStatistics} to evaluate the relative contribution of six groups of predictors---Betti numbers, cell counts, network properties, publication volume, and field and year fixed effects. The results are summarized in Table \ref{table:DominanceAnalysis}. As may be expected, the most informative predictor is the field of publication. However, among the remaining predictors, the Betti numbers (i.e., $\beta_0$, $\beta_1$, $\beta_2$, and $\beta_3$) contribute the most (14.63\%) to the adjusted $R^2$; of note, this contribution is roughly 40\% more than that made by the basic (lower level) network properties and 43\% more than that made by the year fixed effects. 


The relationship we observe between knowledge network topology and the probability of a ``hit'' is robust to alternative model specifications. First, we examined alternative lags between knowledge network topology and hit probability. Table \ref{table:LagRegressions} presents models analogous to those of Table \ref{table:MainRegressions} but with topological properties measured at time $t-1$; the results are similar to those of our main models. Second, for reasons of interpretability and computation, we estimate our statistical models using OLS; however, we also found similar results using nonlinear models (available upon request). Finally, we repeated the statistical analyses discussed above but excluding the APS data, using patents alone. The results of these analyses are shown alongside our main models in the associated regression tables. While there are minor differences across specifications, the overall pattern of results is remarkably similar.

\begin{figure}[t]
\includegraphics[width=\textwidth]{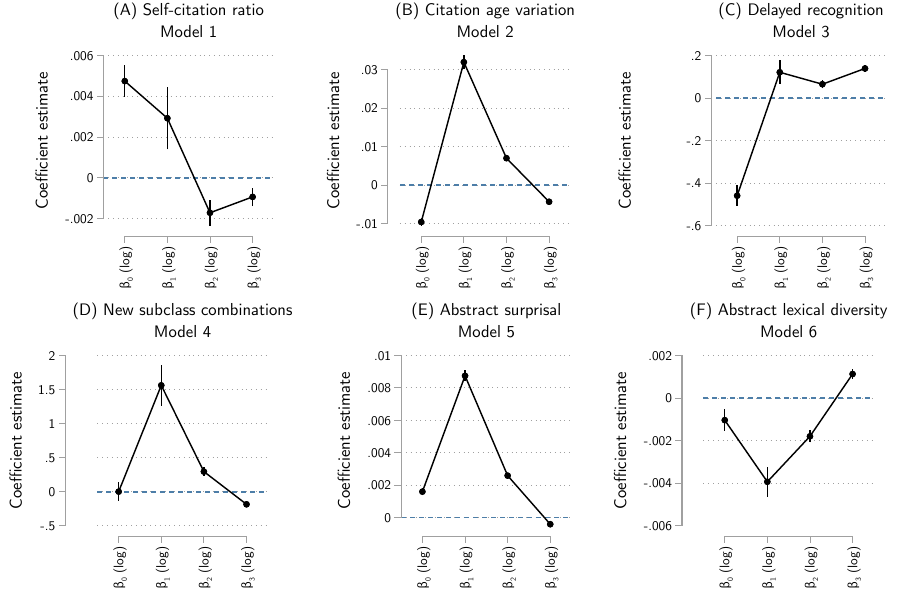}
\caption{\textbf{Coefficient estimates for Betti numbers from Table \ref{table:SupplementalRegressions}}. Model numbers are given above each plot. Error bars represent 95\% confidence intervals. Error bars that span 0 indicate that the corresponding coefficient is not significantly different from 0; coefficients with overlapping error bars may however be significantly different from each other.}
\label{figure:CombinedSupplementalRegressionsCoefficientPlotHoles}
\end{figure}

We conducted several additional analyses (see Table \ref{table:SupplementalRegressions}) to examine the implications of higher-order structure for discovery and invention. Models 1 and 2 of Table \ref{table:SupplementalRegressions} evaluate the relationship between knowledge network topology and search depth. If higher-order structure creates opportunities for new kinds of breakthroughs, then increases in higher-order structure may prompt investigators to comb more deeply through prior work for solutions. We consider two proxies for search depth, (1) the ratio of self-citations to total citations and (2) the variation in the age of prior work cited. Our presumption is that higher values of both the former and latter will reflect lesser and greater search depth, respectively. Figure \ref{figure:CombinedSupplementalRegressionsCoefficientPlotHoles} plots coefficient estimates for the Betti numbers ($\beta_0$, $\beta_1$, $\beta_2$, and $\beta_3$) from these models. The results are consistent with our expectations; at higher dimensions, increases in holes are associated with fewer self citations, while the corresponding pattern for citation age variation is inverted-U shaped. 

Models 4 and 5 evaluate the relationship between knowledge network topology and novelty. Paralleling our thinking on search depth, if higher-order structure creates opportunities for new kinds of breakthroughs, then increases in higher-order structure may prompt investigators to try out ideas that are more distinctive vis-a-vis what has been done before. We consider two proxies for novelty, (1) new subclass combinations and (2) the Jensen-Shannon divergence (i.e., surprisal) of publications (based on the distribution of word frequencies in their abstracts). Once again, we observe results consistent with the idea that up to a point, increasing higher-order structure may be generative of discovery and innovation (see Figure \ref{figure:CombinedSupplementalRegressionsCoefficientPlotHoles} for coefficient plots).

Finally, Models 3 and 6 of Table \ref{table:SupplementalRegressions} evaluate the relationship between knowledge network topology and publication complexity. If higher-order structure creates opportunities for new kinds of breakthroughs, then increases in higher-order structure may prompt investigators to try out more complex ideas. We consider two proxies for complexity, (1) delayed recognition (i.e., publications that are slower to gain citations) and (2) lexical diversity (i.e., more unique words per total word). Our rationale for the former proxy was based on the idea that the significance of more complex publications should be harder to recognize and consequently take longer for future work to use; for the latter proxy, our motivation was based on the idea that describing more complex ideas is likely to require more diverse vocabularies. The results (see Figure \ref{figure:CombinedSupplementalRegressionsCoefficientPlotHoles} for coefficient plots) are consistent with our expectations. Increases in higher dimensional holes are positively associated with delayed recognition; the relationship between holes and lexical diversity is U-shaped, with increases in the highest dimensional holes being associated with increases in lexical diversity. 

\section{Discussion}

For decades, scientific and technological knowledge has developed at an unprecedented pace. Yet observers have questioned whether such progress is sustainable. Recently, network science has emerged as a framework for measuring the structure and dynamics of knowledge, and findings from this work lend credence to concerns of slowing progress. However, current approaches are limited because they overlook the higher-order structure of knowledge, instead focusing on lower-level, dyadic interactions among components. Thus, observations of slowing progress may in part stem from the shifting locus of discovery and invention to higher dimensions, which require new lenses to observe.

We drew on methods from algebraic topology to map the dynamic, higher-order structure of knowledge in science and technology. Our analysis led to several noteworthy findings. First, we documented the emergence of higher-order structure across diverse fields of science and technology. Interestingly, the growth of higher-order structure often coincides with the decline of lower-order structure. We further demonstrated that the emergence of higher-order structure in knowledge networks has historically outpaced the emergence of higher-order structure in collaboration networks, and the topology of knowledge and collaboration are generally diverging over time. The topology of knowledge also tends to be diverging across fields, implying the way in which knowledge is brought together across fields is becoming increasingly heterogeneous.

Second, we observe that topological structure is related to the nature of the science and technology produced. As higher-dimensional holes become more prevalent, publications tend to use words that are more abstract, indicative of developing prior work, and also---at least up to a point---more novel. These observations are consistent with the idea that moderate levels of higher-order structure may be generative of discovery and invention. Our findings of a curvilinear (inverted-U shaped) relationship between further underscore this dual interpretation of the emergence of higher-order structure in the knowledge networks of science and technology.

Finally, we demonstrated that the topology of scientific and technological knowledge encodes information that is not captured by existing measures. Associations between our characterizations of higher-order structure and common, lower-order measures of network properties are typically small. Moreover, the variation in higher-order structure we observe cannot be described by simple small-world or preferential-attachment models. This finding suggests that the patterns we observe are unlikely to be artifacts of the data or underlying network structure. We further demonstrated that the distribution of topological holes in the knowledge network of a field is more predictive of hit publications than a host of other factors, including lower-order network measures. Taken together, these results provide compelling evidence that our measures of higher-order structure capture variation that is not captured using traditional approaches.

Our findings are subject to several limitations. First, while our assessments of knowledge networks in technology were based on all patents granted by the USPTO, our analysis of science was limited to journals published by the APS. Thus, work needs to be done to evaluate whether our findings using the APS data generalize to databases that cover more academic disciplines. Second, although our findings suggest that algebraic topology offers an exciting  lens for the study of knowledge networks, computational constraints currently limit their application to very large databases. Finally, our regression analyses are based on observational data, and therefore our findings on the relationship between higher-order structure and various outcomes should not be interpreted as indicating causal relationships.

Notwithstanding these limitations, our study has several implications. First, our findings suggest the need to rethink and theories of invention. Classical theories describe invention as a process of recombination, in which existing components of knowledge are brought together in novel configurations, typically without attending to the dimensionality of the components brought together. Yet our results suggest that dimensionality of knowledge may be important for shaping both the opportunities for and challenges of recombination. Bringing together similar knowledge components but at different levels of dimensionality may, for example, require distinctive creative processes, resulting in qualitatively distinctive inventions. Second, our results suggest the opportunity for more exploration of models of higher-order structure in other networks of interest in Science of Science. While our primary focus was on knowledge networks, several of our analyses pointed to important changes in the topological structure of collaboration, with higher-order structure becoming more prevalent (though not to the same degree as observed for knowledge networks). Prior work suggests that social network position is an important determinant of individual creativity, with positions that span holes in social structures being particularly valuable \citep{burt2004structural}. While this view is intuitively attractive, empirical work has produced a myriad of conflicting findings \citep{obstfeld2005social, fleming2007collaborative}. Yet prior work has not considered the possibility that the benefits of spanning a structural hole are contingent on its dimensionality. Such a possibility may offer one approach for reconciling these conflicting results.

\bibliographystyle{plainnat}
\bibliography{knowledge_tda}

\begin{thebibliography}{42}
\providecommand{\natexlab}[1]{#1}
\providecommand{\url}[1]{\texttt{#1}}
\expandafter\ifx\csname urlstyle\endcsname\relax
  \providecommand{\doi}[1]{doi: #1}\else
  \providecommand{\doi}{doi: \begingroup \urlstyle{rm}\Url}\fi

\bibitem[Acemoglu et~al.(2016)Acemoglu, Akcigit, and
  Kerr]{acemoglu2016innovation}
Daron Acemoglu, Ufuk Akcigit, and William~R Kerr.
\newblock Innovation network.
\newblock \emph{Proceedings of the National Academy of Sciences}, 113\penalty0
  (41):\penalty0 11483--11488, 2016.

\bibitem[Agrawal et~al.(2016)Agrawal, Goldfarb, and
  Teodoridis]{agrawal2016understanding}
Ajay Agrawal, Avi Goldfarb, and Florenta Teodoridis.
\newblock Understanding the changing structure of scientific inquiry.
\newblock \emph{American Economic Journal: Applied Economics}, 8\penalty0
  (1):\penalty0 100--128, 2016.

\bibitem[Aktas et~al.(2019)Aktas, Akbas, and El~Fatmaoui]{aktas2019persistence}
Mehmet~E Aktas, Esra Akbas, and Ahmed El~Fatmaoui.
\newblock Persistence homology of networks: methods and applications.
\newblock \emph{Applied Network Science}, 4\penalty0 (1):\penalty0 61, 2019.

\bibitem[Arbesman(2011)]{arbesman2011quantifying}
Samuel Arbesman.
\newblock Quantifying the ease of scientific discovery.
\newblock \emph{Scientometrics}, 86\penalty0 (2):\penalty0 245--250, 2011.

\bibitem[Barab{\'a}si and Albert(1999)]{barabasi1999emergence}
Albert-L{\'a}szl{\'o} Barab{\'a}si and R{\'e}ka Albert.
\newblock Emergence of scaling in random networks.
\newblock \emph{Science}, 286\penalty0 (5439):\penalty0 509--512, 1999.

\bibitem[Bauer(2019)]{bauer2019Ripser}
Ulrich Bauer.
\newblock Ripser: efficient computation of vietoris-rips persistence barcodes,
  August 2019.
\newblock Preprint.

\bibitem[Bloom et~al.(2020)Bloom, Jones, Van~Reenen, and Webb]{bloom2020ideas}
Nicholas Bloom, Charles~I Jones, John Van~Reenen, and Michael Webb.
\newblock Are ideas getting harder to find?
\newblock \emph{American Economic Review}, 110\penalty0 (4):\penalty0 1104--44,
  2020.

\bibitem[Brysbaert et~al.(2014)Brysbaert, Warriner, and
  Kuperman]{brysbaert2014concreteness}
Marc Brysbaert, Amy~Beth Warriner, and Victor Kuperman.
\newblock Concreteness ratings for 40 thousand generally known english word
  lemmas.
\newblock \emph{Behavior research methods}, 46\penalty0 (3):\penalty0 904--911,
  2014.

\bibitem[Burt(2004)]{burt2004structural}
Ronald~S Burt.
\newblock Structural holes and good ideas.
\newblock \emph{American journal of sociology}, 110\penalty0 (2):\penalty0
  349--399, 2004.

\bibitem[Carlsson(2009)]{carlsson2009topology}
Gunnar Carlsson.
\newblock Topology and data.
\newblock \emph{Bulletin of the American Mathematical Society}, 46\penalty0
  (2):\penalty0 255--308, 2009.

\bibitem[Christianson et~al.(2020)Christianson, Sizemore~Blevins, and
  Bassett]{christianson2020architecture}
Nicolas~H Christianson, Ann Sizemore~Blevins, and Danielle~S Bassett.
\newblock Architecture and evolution of semantic networks in mathematics texts.
\newblock \emph{Proceedings of the Royal Society A}, 476\penalty0
  (2239):\penalty0 20190741, 2020.

\bibitem[Chu and Evans(2018)]{chu2018too}
Johan~SG Chu and James Evans.
\newblock Too many papers? slowed canonical progress in large fields of
  science.
\newblock \emph{SocArxiv}, 2018.

\bibitem[Cohen-Steiner et~al.(2007)Cohen-Steiner, Edelsbrunner, and
  Harer]{cohen2007stability}
David Cohen-Steiner, Herbert Edelsbrunner, and John Harer.
\newblock Stability of persistence diagrams.
\newblock \emph{Discrete \& computational geometry}, 37\penalty0 (1):\penalty0
  103--120, 2007.

\bibitem[Cowen(2011)]{cowen2011great}
Tyler Cowen.
\newblock \emph{The great stagnation: How America ate all the low-hanging fruit
  of modern history, got sick, and will (eventually) feel better: A Penguin
  eSpecial from Dutton}.
\newblock Penguin, 2011.

\bibitem[Cowen and Southwood(2019)]{cowen2019rate}
Tyler Cowen and Ben Southwood.
\newblock Is the rate of scientific progress slowing down?
\newblock Technical report, George Mason University, 2019.

\bibitem[Dworkin et~al.(2019)Dworkin, Shinohara, and
  Bassett]{dworkin2019emergent}
Jordan~D Dworkin, Russell~T Shinohara, and Danielle~S Bassett.
\newblock The emergent integrated network structure of scientific research.
\newblock \emph{PloS one}, 14\penalty0 (4):\penalty0 e0216146, 2019.

\bibitem[Erd\H{o}s and R\'{e}nyi(1959)]{erdos1959random}
Paul Erd\H{o}s and Alfr\'{e}d R\'{e}nyi.
\newblock On random graphs i.
\newblock \emph{Publ. Math. Debrecen}, 6\penalty0 (290-297):\penalty0 18, 1959.

\bibitem[Fleming(2001)]{fleming2001recombinant}
Lee Fleming.
\newblock Recombinant uncertainty in technological search.
\newblock \emph{Management science}, 47\penalty0 (1):\penalty0 117--132, 2001.

\bibitem[Fleming et~al.(2007)Fleming, Mingo, and
  Chen]{fleming2007collaborative}
Lee Fleming, Santiago Mingo, and David Chen.
\newblock Collaborative brokerage, generative creativity, and creative success.
\newblock \emph{Administrative science quarterly}, 52\penalty0 (3):\penalty0
  443--475, 2007.

\bibitem[Foster et~al.(2015)Foster, Rzhetsky, and Evans]{foster2015tradition}
Jacob~G Foster, Andrey Rzhetsky, and James~A Evans.
\newblock Tradition and innovation in scientists’ research strategies.
\newblock \emph{American Sociological Review}, 80\penalty0 (5):\penalty0
  875--908, 2015.

\bibitem[Ghrist(2008)]{ghrist2008barcodes}
Robert Ghrist.
\newblock Barcodes: the persistent topology of data.
\newblock \emph{Bulletin of the American Mathematical Society}, 45\penalty0
  (1):\penalty0 61--75, 2008.

\bibitem[Gordon(2017)]{gordon2017rise}
Robert~J Gordon.
\newblock \emph{The rise and fall of American growth: The US standard of living
  since the civil war}, volume~70.
\newblock Princeton University Press, 2017.

\bibitem[Hatcher(2002)]{hatcher2002algebraic}
Allen Hatcher.
\newblock \emph{Algebraic Topology}.
\newblock Algebraic Topology. Cambridge University Press, 2002.
\newblock ISBN 9780521795401.

\bibitem[Jones(2009)]{jones2009burden}
Benjamin~F Jones.
\newblock The burden of knowledge and the “death of the renaissance man”:
  Is innovation getting harder?
\newblock \emph{The Review of Economic Studies}, 76\penalty0 (1):\penalty0
  283--317, 2009.

\bibitem[Jones(2010)]{jones2010age}
Benjamin~F Jones.
\newblock Age and great invention.
\newblock \emph{The Review of Economics and Statistics}, 92\penalty0
  (1):\penalty0 1--14, 2010.

\bibitem[Leahey et~al.(2017)Leahey, Beckman, and Stanko]{leahey2017prominent}
Erin Leahey, Christine~M Beckman, and Taryn~L Stanko.
\newblock Prominent but less productive: The impact of interdisciplinarity on
  scientists’ research.
\newblock \emph{Administrative Science Quarterly}, 62\penalty0 (1):\penalty0
  105--139, 2017.

\bibitem[L{\"u}tgehetmann et~al.(2020)L{\"u}tgehetmann, Govc, Smith, and
  Levi]{lutgehetmann2020computing}
Daniel L{\"u}tgehetmann, Dejan Govc, Jason~P Smith, and Ran Levi.
\newblock Computing persistent homology of directed flag complexes.
\newblock \emph{Algorithms}, 13\penalty0 (1):\penalty0 19, 2020.

\bibitem[Milojevi{\'c}(2015)]{milojevic2015quantifying}
Sta{\v{s}}a Milojevi{\'c}.
\newblock Quantifying the cognitive extent of science.
\newblock \emph{Journal of Informetrics}, 9\penalty0 (4):\penalty0 962--973,
  2015.

\bibitem[Mukherjee et~al.(2016)Mukherjee, Uzzi, Jones, and
  Stringer]{mukherjee2016new}
Satyam Mukherjee, Brian Uzzi, Ben Jones, and Michael Stringer.
\newblock A new method for identifying recombinations of existing knowledge
  associated with high-impact innovation.
\newblock \emph{Journal of Product Innovation Management}, 33\penalty0
  (2):\penalty0 224--236, 2016.

\bibitem[Obstfeld(2005)]{obstfeld2005social}
David Obstfeld.
\newblock Social networks, the tertius iungens orientation, and involvement in
  innovation.
\newblock \emph{Administrative science quarterly}, 50\penalty0 (1):\penalty0
  100--130, 2005.

\bibitem[Otter et~al.(2017)Otter, Porter, Tillmann, Grindrod, and
  Harrington]{otter2017roadmap}
Nina Otter, Mason~A Porter, Ulrike Tillmann, Peter Grindrod, and Heather~A
  Harrington.
\newblock A roadmap for the computation of persistent homology.
\newblock \emph{EPJ Data Science}, 6\penalty0 (1):\penalty0 17, 2017.

\bibitem[Pan et~al.(2018)Pan, Petersen, Pammolli, and Fortunato]{pan2018memory}
Raj~K Pan, Alexander~M Petersen, Fabio Pammolli, and Santo Fortunato.
\newblock The memory of science: Inflation, myopia, and the knowledge network.
\newblock \emph{Journal of Informetrics}, 12\penalty0 (3):\penalty0 656--678,
  2018.

\bibitem[Porter and Rafols(2009)]{porter2009science}
Alan Porter and Ismael Rafols.
\newblock Is science becoming more interdisciplinary? measuring and mapping six
  research fields over time.
\newblock \emph{Scientometrics}, 81\penalty0 (3):\penalty0 719--745, 2009.

\bibitem[Rzhetsky et~al.(2015)Rzhetsky, Foster, Foster, and
  Evans]{rzhetsky2015choosing}
Andrey Rzhetsky, Jacob~G Foster, Ian~T Foster, and James~A Evans.
\newblock Choosing experiments to accelerate collective discovery.
\newblock \emph{Proceedings of the National Academy of Sciences}, 112\penalty0
  (47):\penalty0 14569--14574, 2015.

\bibitem[Schulz et~al.(2014)Schulz, Mazloumian, Petersen, Penner, and
  Helbing]{schulz2014exploiting}
Christian Schulz, Amin Mazloumian, Alexander~M Petersen, Orion Penner, and Dirk
  Helbing.
\newblock Exploiting citation networks for large-scale author name
  disambiguation.
\newblock \emph{EPJ Data Science}, 3\penalty0 (1):\penalty0 11, 2014.

\bibitem[Schumpeter(1983)]{schumpeter1983theory}
J.A. Schumpeter.
\newblock \emph{The Theory of Economic Development: An Inquiry Into Profits,
  Capital, Credit, Interest, and the Business Cycle}.
\newblock Economics Third World studies. Transaction Books, 1983.
\newblock ISBN 9780878556984.

\bibitem[Shi et~al.(2015)Shi, Foster, and Evans]{shi2015weaving}
Feng Shi, Jacob~G Foster, and James~A Evans.
\newblock Weaving the fabric of science: Dynamic network models of science's
  unfolding structure.
\newblock \emph{Social Networks}, 43:\penalty0 73--85, 2015.

\bibitem[Uzzi et~al.(2013)Uzzi, Mukherjee, Stringer, and
  Jones]{uzzi2013atypical}
Brian Uzzi, Satyam Mukherjee, Michael Stringer, and Ben Jones.
\newblock Atypical combinations and scientific impact.
\newblock \emph{Science}, 342\penalty0 (6157):\penalty0 468--472, 2013.

\bibitem[Varga(2019)]{varga2019shorter}
Attila Varga.
\newblock Shorter distances between papers over time are due to more
  cross-field references and increased citation rate to higher-impact papers.
\newblock \emph{Proceedings of the National Academy of Sciences}, 116\penalty0
  (44):\penalty0 22094--22099, 2019.

\bibitem[Watts and Strogatz(1998)]{watts1998collective}
Duncan~J Watts and Steven~H Strogatz.
\newblock Collective dynamics of ‘small-world’networks.
\newblock \emph{Nature}, 393\penalty0 (6684):\penalty0 440--442, 1998.

\bibitem[Wu et~al.(2019)Wu, Wang, and Evans]{wu2019large}
Lingfei Wu, Dashun Wang, and James~A Evans.
\newblock Large teams develop and small teams disrupt science and technology.
\newblock \emph{Nature}, 566\penalty0 (7744):\penalty0 378--382, 2019.

\bibitem[Wuchty et~al.(2007)Wuchty, Jones, and Uzzi]{wuchty2007increasing}
Stefan Wuchty, Benjamin~F Jones, and Brian Uzzi.
\newblock The increasing dominance of teams in production of knowledge.
\newblock \emph{Science}, 316\penalty0 (5827):\penalty0 1036--1039, 2007.

\end{thebibliography}

\begin{landscape}
\centering
\begin{table}[!htbp]
\caption{Statistics for USPTO and APS data.}
\label{table:StatisticsUSPTO_APS}

\begin{tabular}{@{}cccccccc@{}}
\toprule
\textbf{Subcategory   Name} & \textbf{Field} & \textbf{Total Articles} & \textbf{Max Articles} & \textbf{Max Article Year} & \textbf{Min Articles} & \textbf{Min Article Year} & \textbf{Mean Articles/Year} \\ \midrule
Agriculture                 & Chemicals      & 18934                   & 725                   & 1989                      & 272                   & 2008                      & 541                         \\
Coating                     & Chemicals      & 51431                   & 2297                  & 2000                      & 778                   & 1979                      & 1469                        \\
Gas                         & Chemicals      & 16401                   & 898                   & 2010                      & 290                   & 1979                      & 469                         \\
Organic compounds           & Chemicals      & 104231                  & 5465                  & 1976                      & 2135                  & 2005                      & 2978                        \\
Resins                      & Chemicals      & 110505                  & 4312                  & 2001                      & 1977                  & 1979                      & 3157                        \\
Communications              & Computers      & 277240                  & 24160                 & 2010                      & 1663                  & 1979                      & 7921                        \\
Hardware and software       & Computers      & 368020                  & 35572                 & 2014                      & 1014                  & 1979                      & 9200                        \\
Computer peripherals        & Computers      & 96753                   & 8713                  & 2010                      & 285                   & 1979                      & 2764                        \\
Information storage         & Computers      & 128085                  & 11313                 & 2010                      & 631                   & 1979                      & 3660                        \\
Business methods            & Computers      & 42137                   & 7803                  & 2010                      & 37                    & 1979                      & 1204                        \\
Drugs                       & Drugs          & 204220                  & 11275                 & 2010                      & 1781                  & 1979                      & 5835                        \\
Surgery                     & Drugs          & 120046                  & 7573                  & 2010                      & 784                   & 1979                      & 3430                        \\
Genetics                    & Drugs          & 27613                   & 1579                  & 2010                      & 227                   & 1978                      & 789                         \\
Electrical devices          & Electrical     & 129274                  & 6861                  & 2010                      & 1490                  & 1979                      & 3694                        \\
Electrical lighting         & Electrical     & 68555                   & 4363                  & 2010                      & 733                   & 1979                      & 1959                        \\
Measuring and testing       & Electrical     & 114823                  & 5848                  & 2010                      & 1488                  & 1979                      & 3281                        \\
Nuclear and x-rays          & Electrical     & 58667                   & 3134                  & 2010                      & 718                   & 1979                      & 1676                        \\
Power systems               & Electrical     & 146737                  & 9616                  & 2010                      & 1709                  & 1979                      & 4192                        \\
Semiconductor devices       & Electrical     & 164486                  & 14813                 & 2010                      & 495                   & 1979                      & 4700                        \\
Material processing         & Mechanical     & 156396                  & 5769                  & 2001                      & 3044                  & 1979                      & 4468                        \\
Metal working               & Mechanical     & 93232                   & 3509                  & 2001                      & 1782                  & 1979                      & 2664                        \\
Motors and engines          & Mechanical     & 125247                  & 5250                  & 2002                      & 1924                  & 1979                      & 3578                        \\
Optics                      & Mechanical     & 60978                   & 3884                  & 2006                      & 588                   & 1982                      & 1742                        \\
Transportation              & Mechanical     & 105559                  & 4886                  & 2010                      & 1567                  & 1979                      & 3016                        \\
APS                         & Physics        & 739004                  & 38054                 & 2010                      & 9362                  & 1980                      & 23839                       \\ \bottomrule
\end{tabular}
\end{table}

\begin{flushleft}
\emph{Notes:} High-level statistics of USPTO and APS computed across the years of interest 1976-2010. Subcategory Name corresponds to the NBER subcategory classification name for the USPTO data. 
\end{flushleft}

\end{landscape}


\pagebreak


{

\thispagestyle{empty}
\pagestyle{empty}

\begin{table}[!htbp]\centering
\begin{threeparttable}
\caption{Most common lemmas in patent abstracts by part of speech and Betti number}
\label{table:AbstractWords}

\footnotesize
\begin{tabularx}{\textwidth}{XXXXXXXX}
\toprule
\multicolumn{8}{c}{Nouns}\\
  \multicolumn{2}{c}{$\beta_0$} & \multicolumn{2}{c}{$\beta_1$} & \multicolumn{2}{c}{$\beta_2$} & \multicolumn{2}{c}{$\beta_3$} \\
\cmidrule(lr){1-2} \cmidrule(lr){3-4} \cmidrule(lr){5-6} \cmidrule(lr){7-8}
      Lemma &        $r$ &      Lemma &        $r$ &          Lemma &        $r$ &          Lemma &        $r$ \\
\cmidrule(lr){1-1} \cmidrule(lr){2-2} \cmidrule(lr){3-3} \cmidrule(lr){4-4} \cmidrule(lr){5-5} \cmidrule(lr){6-6} \cmidrule(lr){7-7} \cmidrule(lr){8-8}
      crank &       0.70 &       form &       0.71 &        process &       0.75 &    degradation &       0.66 \\
      lever &       0.67 &  condition &       0.71 &       property &       0.72 &  functionality &       0.65 \\
 rearwardly &       0.64 &       ring &       0.71 &           step &       0.71 &        process &       0.65 \\
     pulley &       0.64 &    carrier &       0.71 &         method &       0.71 &       property &       0.63 \\
      spool &       0.64 &       case &       0.70 &    degradation &       0.71 &         method &       0.63 \\
       shoe &       0.63 &        use &       0.70 &      substrate &       0.70 &      substrate &       0.63 \\
     clutch &       0.63 &   pressure &       0.69 &  functionality &       0.69 &         medium &       0.63 \\
   sprocket &       0.62 &  reduction &       0.69 &         amount &       0.69 &      multicast &       0.62 \\
 engagement &       0.62 &     degree &       0.68 &         medium &       0.68 &          group &       0.61 \\
      wheel &       0.62 &   strength &       0.68 &      invention &       0.68 &           step &       0.61 \\

\end{tabularx}
\begin{tabularx}{\textwidth}{XXXXXXXX}
\midrule
\multicolumn{8}{c}{Verbs}\\
  \multicolumn{2}{c}{$\beta_0$} & \multicolumn{2}{c}{$\beta_1$} & \multicolumn{2}{c}{$\beta_2$} & \multicolumn{2}{c}{$\beta_3$} \\
\cmidrule(lr){1-2} \cmidrule(lr){3-4} \cmidrule(lr){5-6} \cmidrule(lr){7-8}
      Lemma &        $r$ &         Lemma &        $r$ &      Lemma &        $r$ &      Lemma &        $r$ \\
\cmidrule(lr){1-1} \cmidrule(lr){2-2} \cmidrule(lr){3-3} \cmidrule(lr){4-4} \cmidrule(lr){5-5} \cmidrule(lr){6-6} \cmidrule(lr){7-7} \cmidrule(lr){8-8}
      swing &       0.62 &         carry &       0.74 &    contain &       0.71 &       base &       0.63 \\
      pivot &       0.62 &           say &       0.72 &   describe &       0.71 &   describe &       0.62 \\
    actuate &       0.62 &      comprise &       0.72 &    improve &       0.70 &    enhance &       0.62 \\
 journalled &       0.60 &          have &       0.71 &       base &       0.69 &    contain &       0.61 \\
  disengage &       0.60 &          form &       0.70 &    enhance &       0.69 &    improve &       0.61 \\
      coact &       0.60 &  characterize &       0.70 &        use &       0.68 &        add &       0.61 \\
  journaled &       0.60 &         bring &       0.70 &        add &       0.67 &        use &       0.60 \\
      hinge &       0.59 &       contact &       0.69 &     result &       0.67 &     obtain &       0.60 \\
      brake &       0.59 &      separate &       0.69 &    exhibit &       0.67 &    exhibit &       0.60 \\
       grip &       0.59 &       consist &       0.69 &     obtain &       0.66 &     modify &       0.59 \\

\end{tabularx}
\begin{tabularx}{\textwidth}{XXXXXXXX}
\midrule
\multicolumn{8}{c}{Adjectives}\\
  \multicolumn{2}{c}{$\beta_0$} & \multicolumn{2}{c}{$\beta_1$} & \multicolumn{2}{c}{$\beta_2$} & \multicolumn{2}{c}{$\beta_3$} \\
\cmidrule(lr){1-2} \cmidrule(lr){3-4} \cmidrule(lr){5-6} \cmidrule(lr){7-8}
      Lemma &        $r$ &         Lemma &        $r$ &      Lemma &        $r$ &       Lemma &        $r$ \\
\cmidrule(lr){1-1} \cmidrule(lr){2-2} \cmidrule(lr){3-3} \cmidrule(lr){4-4} \cmidrule(lr){5-5} \cmidrule(lr){6-6} \cmidrule(lr){7-7} \cmidrule(lr){8-8}
 upstanding &       0.62 &           low &       0.72 &    present &       0.70 &     present &       0.63 \\
   slidable &       0.62 &      improved &       0.71 &       less &       0.70 &      active &       0.62 \\
    pivotal &       0.61 &  intermediate &       0.71 &     active &       0.70 &        more &       0.61 \\
    movable &       0.61 &          free &       0.70 &       good &       0.69 &        less &       0.60 \\
  pivotable &       0.60 &          such &       0.70 &      mixed &       0.68 &       mixed &       0.60 \\
 engageable &       0.60 &       further &       0.70 &       more &       0.68 &  functional &       0.60 \\
  shiftable &       0.60 &          same &       0.69 &  excellent &       0.68 &     average &       0.60 \\
  eccentric &       0.60 &      suitable &       0.68 &      least &       0.68 &         non &       0.59 \\
   swinging &       0.60 &         other &       0.68 &        non &       0.68 &   following &       0.59 \\
  elongated &       0.59 &        double &       0.68 &     stable &       0.67 &   excellent &       0.59 \\

\end{tabularx}
\begin{tabularx}{\textwidth}{XXXXXXXX}
\midrule
\multicolumn{8}{c}{Adverbs}\\
    \multicolumn{2}{c}{$\beta_0$} & \multicolumn{2}{c}{$\beta_1$} & \multicolumn{2}{c}{$\beta_2$} & \multicolumn{2}{c}{$\beta_3$} \\
\cmidrule(lr){1-2} \cmidrule(lr){3-4} \cmidrule(lr){5-6} \cmidrule(lr){7-8}
        Lemma &        $r$ &          Lemma &        $r$ &           Lemma &        $r$ &          Lemma &        $r$ \\
\cmidrule(lr){1-1} \cmidrule(lr){2-2} \cmidrule(lr){3-3} \cmidrule(lr){4-4} \cmidrule(lr){5-5} \cmidrule(lr){6-6} \cmidrule(lr){7-7} \cmidrule(lr){8-8}
    pivotally &       0.66 &       together &       0.72 &             e g &       0.71 &            e g &       0.63 \\
    forwardly &       0.65 &        thereof &       0.72 &          highly &       0.70 &         highly &       0.61 \\
  resiliently &       0.64 &   continuously &       0.70 &         wherein &       0.69 &        wherein &       0.59 \\
   adjustably &       0.63 &    essentially &       0.70 &           where &       0.68 &          where &       0.59 \\
     upwardly &       0.62 &        whereby &       0.69 &           least &       0.68 &    efficiently &       0.58 \\
 transversely &       0.61 &  substantially &       0.68 &      optionally &       0.66 &          least &       0.57 \\
    rotatably &       0.61 &     relatively &       0.68 &   significantly &       0.66 &  significantly &       0.57 \\
   lengthwise &       0.60 &       directly &       0.68 &  advantageously &       0.64 &     optionally &       0.57 \\
    drivingly &       0.60 &          first &       0.68 &     efficiently &       0.63 &         herein &       0.57 \\
      rigidly &       0.60 &          where &       0.68 &           about &       0.63 &   specifically &       0.56 \\
\bottomrule
\end{tabularx}

\begin{tablenotes}
\item \emph{Notes:} Parts of speech within USPTO abstracts with highest Spearman correlation with Betti numbers. Note above results are comparable to using OLS and controlling for field. 
\end{tablenotes}

\end{threeparttable}
\end{table}

}%


\pagebreak


\begin{landscape}

{

\footnotesize

\thispagestyle{empty}



\begin{table}[htbp]\centering
\begin{threeparttable}
\def\sym#1{\ifmmode^{#1}\else\(^{#1}\)\fi}
\caption{Summary statistics for variables used in regression analyses}
\label{table:SummaryStatistics}
\begin{tabular}{l*{7}{c}}
\toprule

Variable	&	$N$	&	$N_{unique}$	&	Mean & SD	&	Min	&	Max	& Level of measurement \\ \midrule

 Topological & \\   
 $\beta\textsubscript{0}$ (log) &  4283637 & 198 & 3.30 & 1.31 & 0.00 & 5.95 & Field $\times$ year \\  
 $\beta\textsubscript{1}$ (log) &  4283637 & 755 & 6.17 & 1.14 & 0.00 & 7.91 & Field $\times$ year \\  
 $\beta\textsubscript{2}$ (log) &  4283637 & 472 & 5.04 & 2.07 & 0.00 & 8.22 & Field $\times$ year \\  
 $\beta\textsubscript{3}$ (log) &  4278984 & 216 & 3.13 & 2.50 & 0.00 & 7.87 & Field $\times$ year \\  
 $\Delta\textsubscript{0}$ (log) &  4283637 & 1248 & 8.39 & 1.06 & 0.00 & 9.82 & Field $\times$ year \\  
 $\Delta\textsubscript{1}$ (log) &  4283637 & 1347 & 10.38 & 1.45 & 0.00 & 12.45 & Field $\times$ year \\  
 $\Delta\textsubscript{2}$ (log) &  4283637 & 1361 & 11.66 & 2.00 & 0.00 & 15.94 & Field $\times$ year \\  
 $\Delta\textsubscript{3}$ (log) &  4283637 & 1357 & 12.72 & 2.61 & 0.00 & 19.75 & Field $\times$ year \\  
 $\Delta\textsubscript{4}$ (log) &  4278984 & 1357 & 13.70 & 3.24 & 0.00 & 23.17 & Field $\times$ year \\  
 \midrule   
 Outcomes & \\   
 Hit publication &  4283676 & 2 & 0.05 & 0.22 & 0.00 & 1.00 & Publication \\  
 Citation age variation &  3886365 & 611288 & 0.47 & 0.31 & -143.92 & 82.31 & Publication \\  
 Self-citation ratio &  3892496 & 6156 & 0.08 & 0.18 & 0.00 & 1.00 & Publication \\  
 Delayed recognition &  3896371 & 652 & 1.14 & 10.77 & -1007.00 & 2339.00 & Publication \\  
 New subclass combinations &  3860383 & 812 & 3.49 & 22.73 & 0.00 & 15867.00 & Publication \\  
 Abstract surprisal &  3847898 & 3786985 & 0.30 & 0.06 & 0.10 & 0.97 & Publication \\  
 Abstract lexical diversity &  4116571 & 22422 & 0.54 & 0.13 & 0.06 & 1.00 & Publication \\  
 \midrule   
 Controls & \\   
 Publications (log) &  4283637 & 1188 & 8.29 & 1.32 & 0.00 & 10.09 & Field $\times$ year \\  
 Knowledge network density &  4251206 & 1364 & 0.00 & 0.01 & 0.00 & 1.00 & Field $\times$ year \\  
 Knowledge network clustering &  4251206 & 1360 & 0.30 & 0.10 & 0.15 & 1.00 & Field $\times$ year \\  
 Knowledge network path efficiency &  4283637 & 1365 & 0.23 & 0.08 & 0.00 & 1.00 & Field $\times$ year \\  
 \midrule   
 Fixed effects & \\   
 Year &  4283637 & 42 & -- & -- & -- & -- & Year \\  
 Field &  4283676 & 38 & -- & -- & -- & -- & Field \\

\bottomrule
\end{tabular}

\end{threeparttable}
\end{table}

}%

\end{landscape}


\pagebreak


\begin{landscape}

\thispagestyle{empty}
\pagestyle{empty}

{

\scriptsize

\setlength{\tabcolsep}{2pt}

\renewcommand{\arraystretch}{1}

\begin{table}[htbp]\centering
\begin{threeparttable}
\def\sym#1{\ifmmode^{#1}\else\(^{#1}\)\fi}
\caption{Regressions predicting ``hit'' publications}
\label{table:MainRegressions}
\begin{tabular}{l*{10}{c}}
\toprule

 
                                             &\multicolumn{5}{c}{\shortstack{Sample: USPTO + APS}}                                          &\multicolumn{5}{c}{\shortstack{Sample: USPTO}}                                                \\\cmidrule(lr){2-6}\cmidrule(lr){7-11}
                                             &\multicolumn{1}{c}{(1)}   &\multicolumn{1}{c}{(2)}   &\multicolumn{1}{c}{(3)}   &\multicolumn{1}{c}{(4)}   &\multicolumn{1}{c}{(5)}   &\multicolumn{1}{c}{(6)}   &\multicolumn{1}{c}{(7)}   &\multicolumn{1}{c}{(8)}   &\multicolumn{1}{c}{(9)}   &\multicolumn{1}{c}{(10)}   \\
\midrule
$\beta\textsubscript{0}$ (log)               &        -0.0207***&        -0.0013** &        -0.0053***&                  &        -0.0031***&        -0.0306***&        -0.0053***&        -0.0068***&                  &        -0.0048***\\
                                             &       (0.0004)   &       (0.0004)   &       (0.0004)   &                  &       (0.0005)   &       (0.0004)   &       (0.0005)   &       (0.0005)   &                  &       (0.0005)   \\
$\beta\textsubscript{1}$ (log)               &         0.0072***&         0.0132***&         0.0170***&                  &         0.0163***&         0.0033***&         0.0136***&         0.0137***&                  &         0.0134***\\
                                             &       (0.0003)   &       (0.0005)   &       (0.0005)   &                  &       (0.0005)   &       (0.0004)   &       (0.0005)   &       (0.0005)   &                  &       (0.0005)   \\
$\beta\textsubscript{2}$ (log)               &         0.0060***&         0.0059***&         0.0069***&                  &         0.0074***&         0.0078***&         0.0059***&         0.0066***&                  &         0.0064***\\
                                             &       (0.0001)   &       (0.0002)   &       (0.0002)   &                  &       (0.0002)   &       (0.0001)   &       (0.0002)   &       (0.0002)   &                  &       (0.0002)   \\
$\beta\textsubscript{3}$ (log)               &         0.0106***&         0.0019***&         0.0007***&                  &         0.0015***&         0.0089***&         0.0011***&        -0.0001   &                  &         0.0015***\\
                                             &       (0.0001)   &       (0.0002)   &       (0.0001)   &                  &       (0.0002)   &       (0.0001)   &       (0.0002)   &       (0.0001)   &                  &       (0.0002)   \\
$\Delta\textsubscript{0}$ (log)              &         0.0378***&        -0.0282***&                  &                  &        -0.0427***&         0.0789***&        -0.1045***&                  &                  &        -0.0969***\\
                                             &       (0.0019)   &       (0.0040)   &                  &                  &       (0.0044)   &       (0.0025)   &       (0.0043)   &                  &                  &       (0.0045)   \\
$\Delta\textsubscript{1}$ (log)              &         0.0118** &         0.0518***&                  &                  &         0.0971***&        -0.0022   &         0.1720***&                  &                  &         0.1644***\\
                                             &       (0.0039)   &       (0.0061)   &                  &                  &       (0.0065)   &       (0.0057)   &       (0.0070)   &                  &                  &       (0.0072)   \\
$\Delta\textsubscript{2}$ (log)              &        -0.1695***&        -0.0747***&                  &                  &        -0.1088***&        -0.2262***&        -0.2131***&                  &                  &        -0.2018***\\
                                             &       (0.0051)   &       (0.0062)   &                  &                  &       (0.0066)   &       (0.0077)   &       (0.0082)   &                  &                  &       (0.0081)   \\
$\Delta\textsubscript{3}$ (log)              &         0.1365***&         0.0396***&                  &                  &         0.0662***&         0.2092***&         0.1436***&                  &                  &         0.1353***\\
                                             &       (0.0037)   &       (0.0040)   &                  &                  &       (0.0045)   &       (0.0057)   &       (0.0059)   &                  &                  &       (0.0058)   \\
$\Delta\textsubscript{4}$ (log)              &        -0.0380***&        -0.0070***&                  &                  &        -0.0191***&        -0.0651***&        -0.0407***&                  &                  &        -0.0394***\\
                                             &       (0.0011)   &       (0.0012)   &                  &                  &       (0.0014)   &       (0.0018)   &       (0.0018)   &                  &                  &       (0.0018)   \\
\midrule Year fixed effects                  &             No   &            Yes   &            Yes   &            Yes   &            Yes   &             No   &            Yes   &            Yes   &            Yes   &            Yes   \\
Field fixed effects                          &             No   &            Yes   &            Yes   &            Yes   &            Yes   &             No   &            Yes   &            Yes   &            Yes   &            Yes   \\
Controls                                     &            Yes   &             No   &            Yes   &            Yes   &            Yes   &            Yes   &             No   &            Yes   &            Yes   &            Yes   \\
\midrule
N                                            &        4246553   &        4246553   &        4246553   &        4246553   &        4246553   &        3855730   &        3855730   &        3855730   &        3855730   &        3855730   \\
Adjusted R2                                  &           0.02   &           0.04   &           0.04   &           0.04   &           0.04   &           0.02   &           0.04   &           0.04   &           0.04   &           0.04   \\
\midrule Wald tests for topology predictors  &                  &                  &                  &                  &                  &                  &                  &                  &                  &                  \\
F                                            &        4308.74   &         476.44   &         878.95   &                  &         435.54   &        3936.60   &         540.61   &         716.91   &                  &         400.50   \\
d.f.                                         &           9.00   &           9.00   &           4.00   &                  &           9.00   &           9.00   &           9.00   &           4.00   &                  &           9.00   \\
p-value                                      &           0.00   &           0.00   &           0.00   &                  &           0.00   &           0.00   &           0.00   &           0.00   &                  &           0.00   \\

\bottomrule

\end{tabular}

\begin{tablenotes}
\item \emph{Notes:} Estimates are from ordinary-least-squares regressions (linear probability models). This table evaluates the association between knowledge network topology (measured at the level of the field $\times$ year) and the probability of a ``hit'' publication (patent or paper). Hit publications are defined (using a 0/1 indicator variable) as those that are cited more than 95 percent of all other publications across fields and years, as of 5 years after publication. When indicated, included control variables are measures of the number of publications (logged) and knowledge network density for each field $\times$ year observation. For more details on variables, see Table~\ref{table:SummaryStatistics}. Wald tests reported below each model evaluate whether the included topological predictors significantly improve model fit. Robust standard errors are shown in parentheses; p-values correspond to two-tailed tests. 
\item {+}p<0.1; {*}p<0.05; {**}p<0.01; {***}p<0.001
\end{tablenotes}

\end{threeparttable}
\end{table}

}%

\end{landscape}


\pagebreak


\thispagestyle{empty}
\pagestyle{empty}

{

\scriptsize

\setlength{\tabcolsep}{2pt}

\renewcommand{\arraystretch}{1}

\begin{table}[htbp]\centering
\begin{threeparttable}
\caption{Decomposition of adjusted $R^2$ from regression models predicting ``hit'' publications}
\label{table:DominanceAnalysis}
\begin{tabular}{lccc}
\toprule

\multicolumn{4}{c}{(a) Sample: USPTO + APS} \\ \midrule

Variable set & \multicolumn{3}{c}{Contribution to adjusted $R^2$} \\ \cmidrule(lr){1-1} \cmidrule(lr){2-4}
& Raw & Percent & Rank \\  

$\beta\textsubscript{0}$ (log), $\beta\textsubscript{1}$ (log), $\beta\textsubscript{2}$ (log), $\beta\textsubscript{3}$ (log)	&	0.0058 &	14.63 &	2 \\
$\Delta\textsubscript{0}$ (log), $\Delta\textsubscript{1}$ (log), $\Delta\textsubscript{2}$ (log), $\Delta\textsubscript{3}$ (log), $\Delta\textsubscript{4}$ (log)	&	0.0050 &	12.69 &	3 \\
Knowledge network density, Knowledge network clustering, Knowledge network path efficiency	&	0.0039 &	9.78 &	4 \\ 
Publications (log)	&	0.0008 &	1.95 &	6 \\
Year fixed effects	&	0.0037 &	9.42 &	5 \\
Field fixed effects	&	0.0205 &	51.54 &	1 \\ \midrule

N &                              4246553 \\
Overall adjusted $R^2$     &                 0.0397 \\ 
\midrule

\multicolumn{4}{c}{(b) Sample: USPTO} \\ \midrule
Variable set & \multicolumn{3}{c}{Contribution to adjusted $R^2$} \\ \cmidrule(lr){1-1} \cmidrule(lr){2-4}
& Raw & Percent & Rank \\ 

$\beta\textsubscript{0}$ (log), $\beta\textsubscript{1}$ (log), $\beta\textsubscript{2}$ (log), $\beta\textsubscript{3}$ (log)       &         0.0051   &   13.53     &       2 \\
$\Delta\textsubscript{0}$ (log), $\Delta\textsubscript{1}$ (log), $\Delta\textsubscript{2}$ (log), $\Delta\textsubscript{3}$ (log), $\Delta\textsubscript{4}$ (log)       &         0.0032   &   8.48     &       4 \\
Knowledge network density, Knowledge network clustering, Knowledge network path efficiency      &         0.0028   &   7.48     &       5 \\
Publications (log)       &         0.0003   &   0.85     &       6 \\
Year fixed effects       &         0.0048   &   12.81     &       3 \\
Field fixed effects       &         0.0213   &   56.84     &       1 \\ \midrule

N &                     3855730 \\
Overall adjusted $R^2$     &                 0.0376 \\

\bottomrule

\end{tabular}

\begin{tablenotes}
\item \emph{Notes:} See Models 5 and 10 of Table~\ref{table:MainRegressions} for the full regressions underpinning the decompositions in panels (a) and (b) above, respectively.
\end{tablenotes}

\end{threeparttable}
\end{table}

}%


\pagebreak


\begin{landscape}

\thispagestyle{empty}
\pagestyle{empty}

{

\scriptsize

\setlength{\tabcolsep}{2pt}

\renewcommand{\arraystretch}{1}

\begin{table}[htbp]\centering
\begin{threeparttable}
\def\sym#1{\ifmmode^{#1}\else\(^{#1}\)\fi}
\caption{Regressions predicting search depth, novelty, and complexity of publications}
\label{table:SupplementalRegressions}
\begin{tabular}{l*{10}{c}}
\toprule

 
                                             &\multicolumn{3}{c}{\shortstack{Sample: USPTO + APS}}    &\multicolumn{3}{c}{\shortstack{Sample: USPTO}}          \\\cmidrule(lr){2-4}\cmidrule(lr){5-7}
                                             &\multicolumn{1}{c}{(1)}&\multicolumn{1}{c}{(2)}&\multicolumn{1}{c}{(3)}&\multicolumn{1}{c}{(4)}&\multicolumn{1}{c}{(5)}&\multicolumn{1}{c}{(6)}\\
                                             &\multicolumn{1}{c}{\shortstack{Self-citation \\ ratio \\ (Proxy for \\ search depth)}}&\multicolumn{1}{c}{\shortstack{Citation age \\ variation \\ (Proxy for \\ search depth)}}&\multicolumn{1}{c}{\shortstack{Delayed \\ recognition \\ (Proxy for \\ novelty)}}&\multicolumn{1}{c}{\shortstack{New subclass \\ combinations  \\ (Proxy for \\ novelty)}}&\multicolumn{1}{c}{\shortstack{Abstract \\ surprisal \\ (Proxy for \\ novelty)}}&\multicolumn{1}{c}{\shortstack{Abstract \\ lexical diversity  \\ (Proxy for \\ complexity)}}\\
\midrule
$\beta\textsubscript{0}$ (log)               &         0.0047***&        -0.0095***&        -0.4582***&         0.0026   &         0.0016***&        -0.0010***\\
                                             &       (0.0004)   &       (0.0005)   &       (0.0244)   &       (0.0690)   &       (0.0001)   &       (0.0003)   \\
$\beta\textsubscript{1}$ (log)               &         0.0029***&         0.0319***&         0.1210***&         1.5614***&         0.0087***&        -0.0039***\\
                                             &       (0.0008)   &       (0.0009)   &       (0.0285)   &       (0.1518)   &       (0.0002)   &       (0.0004)   \\
$\beta\textsubscript{2}$ (log)               &        -0.0017***&         0.0070***&         0.0650***&         0.2964***&         0.0026***&        -0.0018***\\
                                             &       (0.0003)   &       (0.0004)   &       (0.0099)   &       (0.0330)   &       (0.0001)   &       (0.0001)   \\
$\beta\textsubscript{3}$ (log)               &        -0.0009***&        -0.0043***&         0.1389***&        -0.1834***&        -0.0004***&         0.0011***\\
                                             &       (0.0002)   &       (0.0003)   &       (0.0084)   &       (0.0209)   &       (0.0001)   &       (0.0001)   \\
$\Delta\textsubscript{0}$ (log)              &         0.0144*  &         0.0001   &         1.6051***&        -3.0818+  &        -0.0159***&         0.1028***\\
                                             &       (0.0057)   &       (0.0060)   &       (0.2050)   &       (1.6305)   &       (0.0015)   &       (0.0031)   \\
$\Delta\textsubscript{1}$ (log)              &        -0.0126   &        -0.1465***&        -4.5364***&        -9.1558+  &        -0.0120***&        -0.1031***\\
                                             &       (0.0094)   &       (0.0092)   &       (0.2940)   &       (4.7986)   &       (0.0025)   &       (0.0052)   \\
$\Delta\textsubscript{2}$ (log)              &         0.0360***&         0.0696***&         2.0355***&        13.9286*  &         0.0119***&         0.0288***\\
                                             &       (0.0098)   &       (0.0096)   &       (0.2653)   &       (5.9577)   &       (0.0027)   &       (0.0058)   \\
$\Delta\textsubscript{3}$ (log)              &        -0.0284***&        -0.0064   &        -0.4039*  &        -7.4918*  &        -0.0036+  &         0.0011   \\
                                             &       (0.0063)   &       (0.0065)   &       (0.1620)   &       (3.6221)   &       (0.0019)   &       (0.0040)   \\
$\Delta\textsubscript{4}$ (log)              &         0.0103***&         0.0000   &        -0.1336** &         1.8002*  &         0.0016** &        -0.0022+  \\
                                             &       (0.0019)   &       (0.0020)   &       (0.0503)   &       (0.9110)   &       (0.0006)   &       (0.0012)   \\
\midrule Year fixed effects                  &            Yes   &            Yes   &            Yes   &            Yes   &            Yes   &            Yes   \\
Field fixed effects                          &            Yes   &            Yes   &            Yes   &            Yes   &            Yes   &            Yes   \\
Controls                                     &            Yes   &            Yes   &            Yes   &            Yes   &            Yes   &            Yes   \\
\midrule
N                                            &        2956056   &        3852420   &        3859287   &        3855730   &        3843247   &        3843249   \\
Adjusted R2                                  &           0.05   &           0.21   &           0.06   &           0.01   &           0.19   &           0.16   \\

\bottomrule

\end{tabular}

\begin{tablenotes}
\item \emph{Notes:} Estimates are from ordinary-least-squares regressions (linear probability models). This table evaluates the association between knowledge network topology (measured at the level of the field $\times$ year) and proxies for the search depth, novelty, and complexity of publications (patents and papers). Included control variables are measures of the number of publications (logged) and knowledge network density for each field $\times$ year observation. Model 1 is conditional on the possibility of self-citation (i.e., at least one member of the author team must have published previously). For more details on variables, see Table~\ref{table:SummaryStatistics}. Robust standard errors are shown in parentheses; p-values correspond to two-tailed tests. 
\item {+}p<0.1; {*}p<0.05; {**}p<0.01; {***}p<0.001
\end{tablenotes}

\end{threeparttable}
\end{table}

}%

\end{landscape}


\pagebreak

\beginsupplement

\section{Supplementary Materials}

\subsection{Methods}

\subsubsection{Persistence Computation}\label{sec:persistence_computation}

For all networks, we computed persistent homology using the Flagser package \citep{lutgehetmann2020computing}. Although originally designed to compute persistent homology of \emph{directed} flag complexes, the package provides good performance in the computation of persistent homology on undirected flag complexes as well. Flagser is built on top of Ripser \citep{bauer2019Ripser}, a high-performance C++ package for computing Vietoris-Rips persistence, and provides high-level improvements to memory management, filtration customization, approximations, and input/output. Computing persistent homology is a high-complexity operation in terms of both computation and memory. The computational complexity of persistent homology is at worst cubic in the number of cells in a particular dimension and is believed to be on the order of matrix multiplication, as the persistence calculation is intrinsically a matrix decomposition of boundary matrices \citep{otter2017roadmap}. Unfortunately, the size of these boundary matrices scales combinatorially with computed homology dimension $k$. For the flag complex of a graph, the number of $k$-cells is the number of $(k-1)$-cliques in the graph. For dense graphs, this number grows combinatorially for increasing $k$ and is at worst $2^n$ for complete graphs on $n$ vertices. Therefore in practice, the computation of persistent homology is generally restricted to the first few dimensions of the complex. We computed persistent homology up to $k=4$ for many of the knowledge networks, although due to computational limitations, only reached $k=3$ for some of the larger, denser networks. For the Drugs NBER subcategory in year 1989, we imputed $\beta_3$ as the average of $\beta_3$ in 1988 and 1990 as the size of the network in this year was intractable. For all large knowledge networks, persistence computations were run on a node using multiple AMD EPYC 7702 processors and 2TB RAM hosted by the Minnesota Supercomputing Institute. Computations on smaller knowledge networks were run on a multi-core Intel Xeon CPU E5-2695 machine with 96GB RAM.

\subsection{Results}

\subsubsection{Additional Knowledge Networks}\label{sec:additional_knowledge_networks}

To help determine whether the topological dynamics we observe using the USPTO and APS data capture distinctive properties of scientific and technological knowledge production or whether similar dynamics would be observed when tracking the growth of other large, two-mode knowledge networks, we computed persistent homology using data from several Stack Exchange sites. Stack Exchange is a network of question-and-answer websites. While perhaps most famous for Stack Overflow, a programming question-and-answer community, Stack Exchange encompasses a diverse range of topics, from cooking (``Seasoned Advice'') and gaming (``Arqade'') to server (``ServerFault'') and database administration (``Database Administrators''). We focus on five large and representative communities---Mathematics, MathOverflow, Physics, Statistical Analysis, and Theoretical Computer Science---from the set of Stack Exchange sites on science topics. For our purposes, Stack Exchange sites are particularly attractive because, similar to the USPTO and APS data, each community includes a well-defined knowledge categorization system, used to tag questions by topic. 

Stack Exchange creates regular snapshots of network websites and makes them freely available for download as database dumps. For our study, we downloaded the snapshot from September 2019, which was the most recent one available at the time of collection. After downloading, we extracted the archives for the five sites of interest. For each site, we then created a two mode edge list, consisting of questions and their associated tags, which was annotated with the date of posting. We subsequently projected this two mode edge list to a one mode representation, after which we were left with a knowledge network consisting of tags that were connected if they appeared together on one or more questions. Similar to our process for the USPTO and APS data, we weight edges based on the number of questions on which the incident tags co-occur. Given the much higher frequency of new questions over time compared to new scientific and technological papers and the much shorter time series (the oldest site in our data, MathOverflow, had been around for only 10 years at the time of data collection), we defined the knowledge networks for the Stack Exchange data on a monthly timescale. 

\begin{figure}[htbp]
\includegraphics[width=\textwidth]{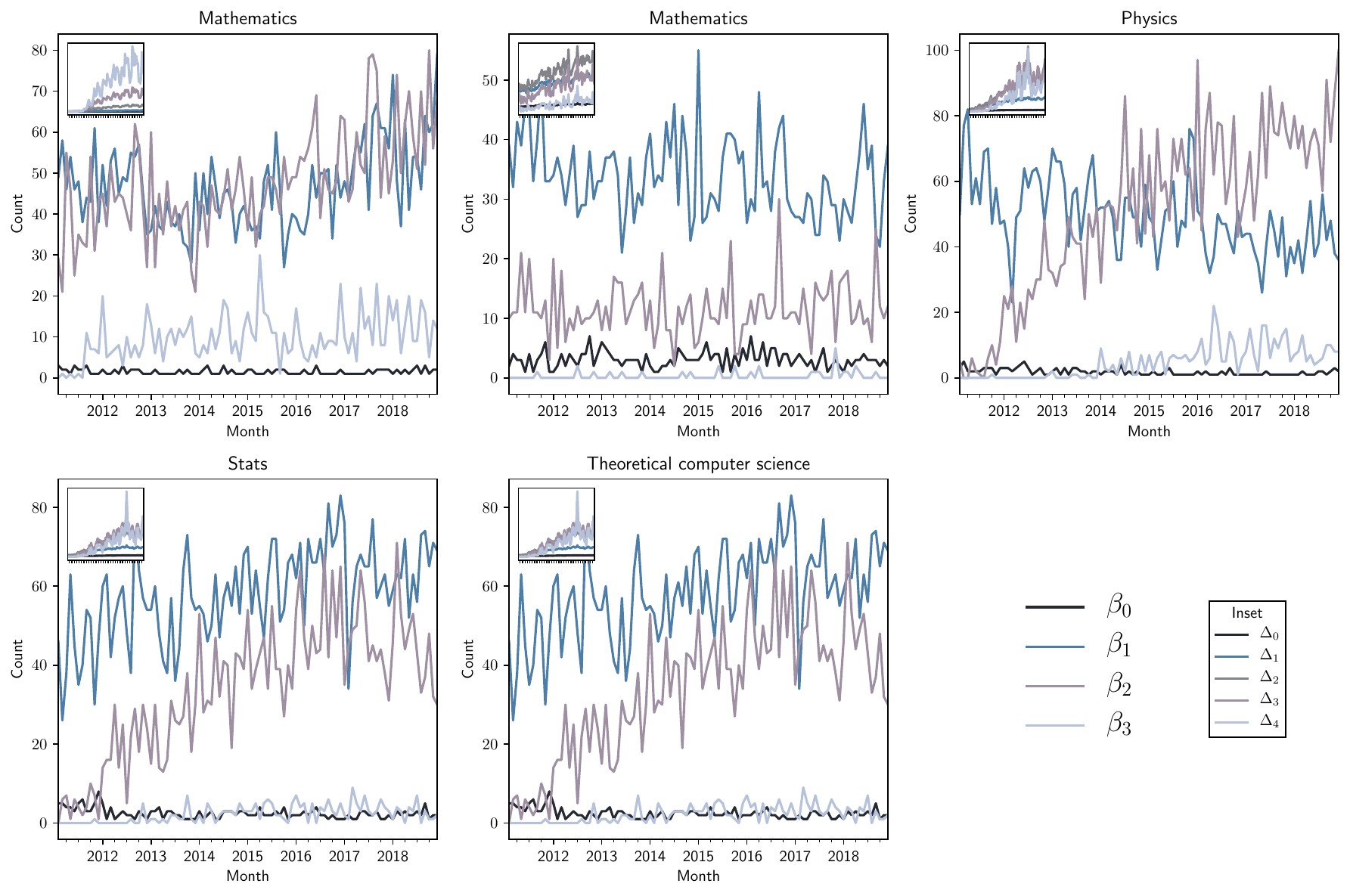}
\caption{\textbf{Knowledge network topology over time for Stack Exchange sites (monthly).} The main plots track counts of Betti numbers, while the inset plots track cell counts. The main plots track counts of Betti numbers, while the inset plots track cell counts. All y-axes are reported on a log scale.}
\label{figure:BettiStackExchangeDynamicsKnowledgeMo}
\end{figure}

Given this weighted network representation of tags connected by questions, we follow exactly the construction described in Section \ref{sec:knowledge_networks}, where now $y \in [01/01/2012,12/01/2019]$ and $c$ is one of the ``Mathematics'', ``MathOverflow'', ``Physics'', ``Statistical Analysis'', or ``Theoretical Computer Science'' sites. We compute the homology of these networks for each site-year combination as was done with the USPTO and APS knowledge networks (Section \ref{sec:persistence_computation}). The distribution of Betti numbers over time is presented in Figure \ref{figure:BettiStackExchangeDynamicsKnowledgeMo}. 

\subsubsection{Linguistic Abstraction}

To begin, we pulled the full text of abstracts for all patents in our data (abstracts were not available for the APS data at the time of analysis), which we then processed using the Natural Language Toolkit and spaCy Python package. Each abstract was split into a list of tokens, and each token was assigned a part-of-speech tag. Separately for each part of speech (we focus on nouns, verbs, adjectives, and adverbs), tokens were grouped by inflected forms into common lemmas. To allow further grouping, we converted all lemmas (again by part of speech) to lower case, removed extraneous punctuation ('\}'!([\$):\%];,\&`?.\{~-\@"\#\_'), and ensured normalization of white space. For each abstract and part of speech, we then dropped duplicate tokens, so that the resulting data consisted of patent $\times$ part of speech $\times$ token tuples. We then aggregated these tuples into a balanced subfield $\times$ year $\times$ part of speech $\times$ token panel and counted the number of patents granted in each subfield $\times$ year using each part of speech $\times$ token in their abstracts. Finally, we joined this panel to our subfield $\times$ year data on Betti numbers by dimension.

\subsubsection{Comparison to Random Networks}\label{sec:random_comparison}

Random graphs play a central role in the study of network complexity, as randomness combined with simple rules of graph construction have been shown to produce complex networks with non-trivial structural features. We were interested in the extent to which the high order structure of knowledge networks may be captured by random graph models and find that, for three popular random graph models, the homological structure of knowledge networks cannot be explained by random models. 

We compared the Betti number distribution across five dimensions of the knowledge networks to the Erd\H{o}s-R\'{e}nyi (ER) \citep{erdos1959random}, Barab\'{a}si-Albert (BA) \citep{barabasi1999emergence}, and Watts-Strogatz (WS) \citep{watts1998collective} random graph models. For each knowledge network (each subcategory-year combination), we constructed 10 random graphs with the same number of nodes and approximately equal number of edges (exactly equal for  BA and WS models). Because the WS model has two free parameters, we constructed 10 WS graphs across 10 linearly-spaced choices of the rewiring probability parameter in range [0.01,0.99]. We computed the homology of each randomly-seeded graph and compared the Betti number distributions of the random graphs to the reference knowledge network. 

For each dimension and each random network, a one-sample t-test was used to compare the sampled mean Betti numbers of the random model across 10 random seeds to the knowledge network's Betti number. Of the 1,608 knowledge networks, no random graph model was statistically equivalent in the Betti distribution across all dimensions. Only the Biotechnology subcategory in 2007 was statistically indistinguishable to the WS model in three dimensions ($\beta_2$, $\beta_3$, and $\beta_4$), but these Betti numbers were all trivial. Only 188 knowledge networks contained at least one dimension that was statistically equivalent to a random model with equivalent node and edge distributions. Clearly, the homology of knowledge networks cannot be explained by ER, BA, or WS random models. 

\begin{figure}[htbp]
\includegraphics[width=\textwidth]{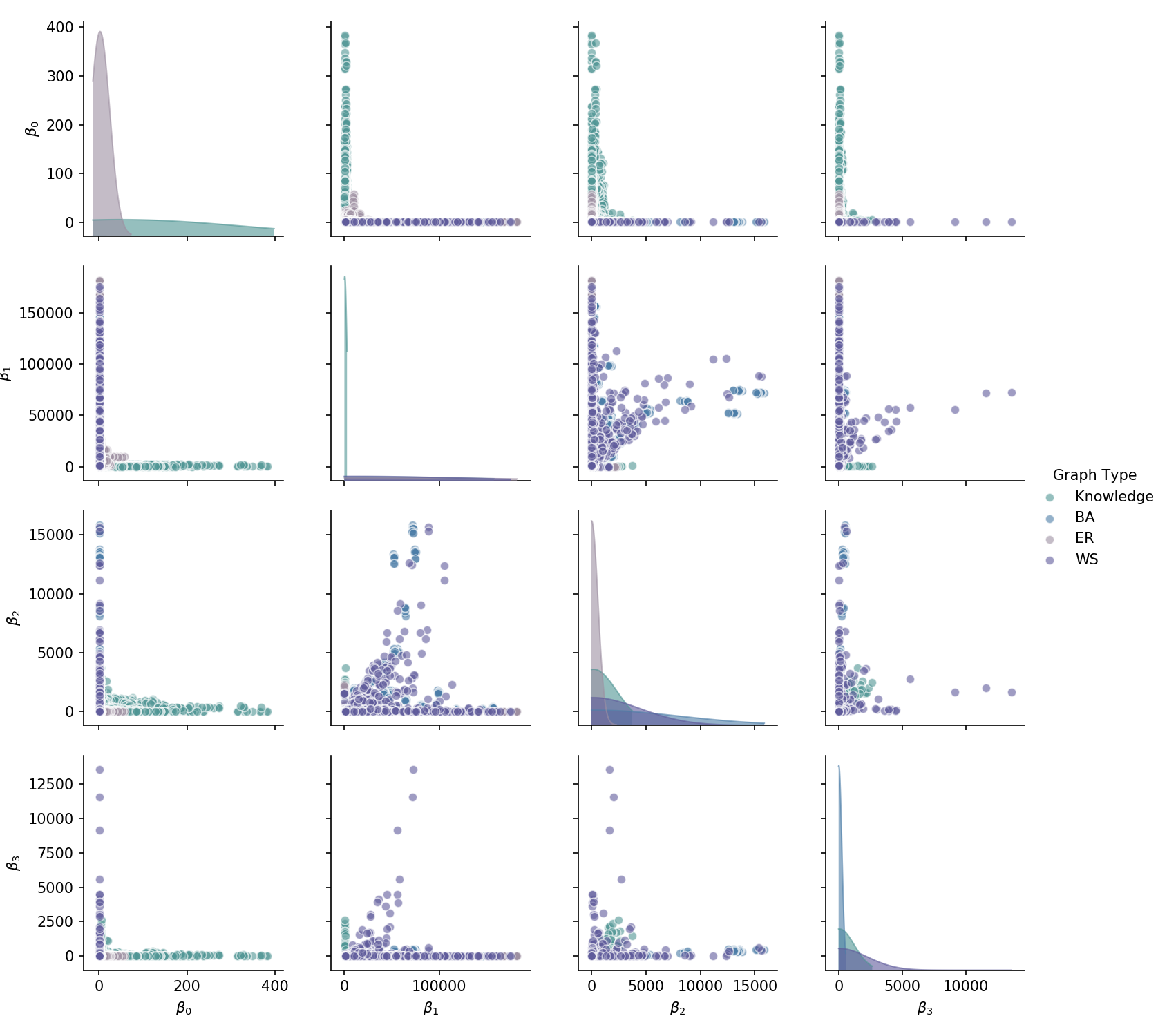}
\caption{\textbf{Distribution differences to topology of random networks.} Ten random instantiations of Erdos-Renyi (ER), Barabasi-Albert (BA), and Watts-Strogatz (WS) random graphs were created to match the edge and node distribution for every year of each subcategory of the knowledge network. For the WS model, 10 linearly-spaced values of rewiring value $p$ were chosen for each random initialization, and here only one random initialization is shown for each $p$. The diagonals depict a kernel density estimate for the distributions of the Betti number in each dimension. The off-diagonals depict cross-sections of the distribution across pairs of dimensions.}
\label{figure:RandomNetworkKnowledgeMatch}
\end{figure}

Figure \ref{figure:RandomNetworkKnowledgeMatch} shows the distribution of Betti numbers across the random and the knowledge networks. The diagonal depicts a kernel density estimate of the distributions in each dimension while the off-diagonal cells show pairwise cross-sections of Betti numbers. The knowledge networks contain more connected components than the random graph models. In fact, ER is the only model that can produce sizeable graphs with  $\beta_0 > 1$, as WS can produce disconnected graphs only if the number of nodes is small enough. This difference in $\beta_0$ is not surprising given BA is constructed by connecting new nodes to a main component and WS assumes full neighborhood connectivity and later re-wires. The most surprising difference between the knowledge and random networks is in the first dimension. The size of $\beta_1$ in all of the random models is significantly larger than in the corresponding knowledge networks. In short, knowledge networks close 1-dimensional holes at a much higher frequency than would be expected in a randomly distributed model with equivalent node and edge distributions. In the higher dimensions, distributional overlap is closer, but the BA and WS models still show significantly longer tails while the ER model only produces non-trivial $\beta_2$ and only for dense networks with few nodes. 

This comparison to random models shows that the homological structure of knowledge networks cannot be described by simple construction rules. The homology of knowledge networks follows neither a preferential attachment (BA) nor small-world (WS) model. Instead, knowledge networks show a much smoother distribution of Betti numbers across dimensions wherein the number of holes in each dimension increases up to $\beta_1$ or $\beta_2$ and then decreases in higher dimension. This is in contrast to the BA and WS models which in most parameter combinations show large jumps between $\beta_i$ and $\beta_{i+1}$. As well, the relative lack of 1-dimensional holes in knowledge networks implies some type of higher-order preferential attachment with respect to triplets of knowledge areas in the network. Unpacking this observation offers an exciting route for future investigation.

\begin{landscape}

\thispagestyle{empty}
\pagestyle{empty}

{

\scriptsize

\setlength{\tabcolsep}{2pt}

\renewcommand{\arraystretch}{1}

\begin{table}[htbp]\centering
\begin{threeparttable}
\def\sym#1{\ifmmode^{#1}\else\(^{#1}\)\fi}
\caption{Regressions predicting ``hit'' publications using lagged topological measures}
\label{table:LagRegressions}
\begin{tabular}{l*{10}{c}}
\toprule

 
                                             &\multicolumn{5}{c}{\shortstack{Sample: USPTO + APS}}                                          &\multicolumn{5}{c}{\shortstack{Sample: USPTO}}                                                \\\cmidrule(lr){2-6}\cmidrule(lr){7-11}
                                             &\multicolumn{1}{c}{(1)}   &\multicolumn{1}{c}{(2)}   &\multicolumn{1}{c}{(3)}   &\multicolumn{1}{c}{(4)}   &\multicolumn{1}{c}{(5)}   &\multicolumn{1}{c}{(6)}   &\multicolumn{1}{c}{(7)}   &\multicolumn{1}{c}{(8)}   &\multicolumn{1}{c}{(9)}   &\multicolumn{1}{c}{(10)}   \\
\midrule
$\beta\textsubscript{0}$ (log, t-1)          &        -0.0119***&        -0.0006   &        -0.0014***&                  &        -0.0006   &        -0.0211***&        -0.0056***&        -0.0043***&                  &        -0.0038***\\
                                             &       (0.0003)   &       (0.0004)   &       (0.0004)   &                  &       (0.0004)   &       (0.0004)   &       (0.0005)   &       (0.0005)   &                  &       (0.0005)   \\
$\beta\textsubscript{1}$ (log, t-1)          &        -0.0021***&         0.0137***&         0.0104***&                  &         0.0116***&         0.0018***&         0.0160***&         0.0148***&                  &         0.0115***\\
                                             &       (0.0003)   &       (0.0005)   &       (0.0004)   &                  &       (0.0005)   &       (0.0004)   &       (0.0005)   &       (0.0005)   &                  &       (0.0005)   \\
$\beta\textsubscript{2}$ (log, t-1)          &         0.0060***&         0.0059***&         0.0055***&                  &         0.0054***&         0.0089***&         0.0058***&         0.0046***&                  &         0.0051***\\
                                             &       (0.0001)   &       (0.0002)   &       (0.0002)   &                  &       (0.0002)   &       (0.0001)   &       (0.0002)   &       (0.0002)   &                  &       (0.0002)   \\
$\beta\textsubscript{3}$ (log, t-1)          &         0.0089***&         0.0015***&        -0.0003*  &                  &         0.0027***&         0.0095***&         0.0010***&         0.0003+  &                  &         0.0020***\\
                                             &       (0.0001)   &       (0.0002)   &       (0.0001)   &                  &       (0.0002)   &       (0.0001)   &       (0.0002)   &       (0.0002)   &                  &       (0.0002)   \\
$\Delta\textsubscript{0}$ (log, t-1)         &         0.0312***&        -0.0163***&                  &                  &        -0.0067+  &         0.0700***&        -0.1111***&                  &                  &        -0.0731***\\
                                             &       (0.0018)   &       (0.0037)   &                  &                  &       (0.0039)   &       (0.0025)   &       (0.0045)   &                  &                  &       (0.0047)   \\
$\Delta\textsubscript{1}$ (log, t-1)         &         0.0865***&         0.0486***&                  &                  &         0.0432***&         0.0201***&         0.1879***&                  &                  &         0.1750***\\
                                             &       (0.0039)   &       (0.0061)   &                  &                  &       (0.0061)   &       (0.0057)   &       (0.0073)   &                  &                  &       (0.0073)   \\
$\Delta\textsubscript{2}$ (log, t-1)         &        -0.2717***&        -0.0712***&                  &                  &        -0.0690***&        -0.2737***&        -0.2324***&                  &                  &        -0.2389***\\
                                             &       (0.0051)   &       (0.0061)   &                  &                  &       (0.0062)   &       (0.0078)   &       (0.0085)   &                  &                  &       (0.0085)   \\
$\Delta\textsubscript{3}$ (log, t-1)         &         0.1883***&         0.0284***&                  &                  &         0.0271***&         0.2337***&         0.1512***&                  &                  &         0.1556***\\
                                             &       (0.0035)   &       (0.0038)   &                  &                  &       (0.0040)   &       (0.0059)   &       (0.0061)   &                  &                  &       (0.0061)   \\
$\Delta\textsubscript{4}$ (log, t-1)         &        -0.0484***&        -0.0015   &                  &                  &        -0.0016   &        -0.0702***&        -0.0417***&                  &                  &        -0.0428***\\
                                             &       (0.0011)   &       (0.0011)   &                  &                  &       (0.0012)   &       (0.0018)   &       (0.0019)   &                  &                  &       (0.0019)   \\
\midrule Year fixed effects                  &             No   &            Yes   &            Yes   &            Yes   &            Yes   &             No   &            Yes   &            Yes   &            Yes   &            Yes   \\
Field fixed effects                          &             No   &            Yes   &            Yes   &            Yes   &            Yes   &             No   &            Yes   &            Yes   &            Yes   &            Yes   \\
Controls                                     &            Yes   &             No   &            Yes   &            Yes   &            Yes   &            Yes   &             No   &            Yes   &            Yes   &            Yes   \\
\midrule
N                                            &        4176784   &        4176784   &        4176784   &        4246553   &        4176784   &        3785961   &        3785961   &        3785961   &        3855730   &        3785961   \\
Adjusted R2                                  &           0.02   &           0.04   &           0.04   &           0.04   &           0.04   &           0.02   &           0.04   &           0.04   &           0.04   &           0.04   \\

\bottomrule

\end{tabular}

\begin{tablenotes}
\item \emph{Notes:} Estimates are from ordinary-least-squares regressions (linear probability models). This table evaluates the association between knowledge network topology (measured at the level of the field $\times$ year) and the probability of a ``hit'' publication (patent or paper) using lagged (1 year) topological measures. Hit publications are defined (using a 0/1 indicator variable) as those that are cited more than 95 percent of all other publications across fields and years, as of 5 years after publication. When indicated, included control variables are measures of the number of publications (logged) and knowledge network density for each field $\times$ year observation. For more details on variables, see Table~\ref{table:SummaryStatistics}. Robust standard errors are shown in parentheses; p-values correspond to two-tailed tests. 
\item {+}p<0.1; {*}p<0.05; {**}p<0.01; {***}p<0.001
\end{tablenotes}

\end{threeparttable}
\end{table}

}%

\end{landscape}


\end{document}